\documentclass[fleqn,usenatbib]{mnras}

\usepackage{newtxtext,newtxmath}
\usepackage[T1]{fontenc}
\usepackage{ae,aecompl}

\usepackage{graphicx}
\usepackage{amsmath}	
\usepackage{amssymb}

\usepackage{lscape}
\usepackage{romannum}
\usepackage{booktabs, caption, fixltx2e}
\usepackage[flushleft]{threeparttable}
\usepackage{paralist}

\newcommand{\Msun}{\rm{M}_\odot}
\newcommand{\Mhe}{\rm{M}_{\alpha}}
\newcommand{\Mco}{\rm{M}_{CO}}

\newcommand{\fcbm}{f_{\rm CBM}}
\newcommand{\alphasc}{\alpha_{\rm sc}}
\newcommand{\grada}{\nabla_{\rm ad}}
\newcommand{\gradr}{\nabla_{\rm rad}}

\title[Importance of Convective Uncertainties]{Relative Importance of Convective Uncertainties in Massive Stars}

\author[Kaiser et al.]{
Etienne A. Kaiser$^{1,6}$\thanks{E-mail: e.kaiser@keele.ac.uk}\href{https://orcid.org/0000-0001-7237-2960}{\includegraphics[scale=0.6]{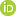}},
Raphael Hirschi$^{1,2,6}$,
W. David Arnett$^{3}$,
Cyril Georgy$^{4}$,\newauthor
Laura J. A.  Scott$^{1}$ and
Andrea Cristini$^{1,5}$
\\
$^{1}$ Keele University, Keele, Staffordshire ST5 5BG, UK\\
$^{2}$ Institute for the Physics and Mathematice of the Universe (WPI), University of Tokyo, 5-1-5 Kashiwanoha, Kashiwa 277-8583, Japan\\
$^{3}$ Steward Observatory, University of Arizona, 933 N. Cherry Avenue, Tucson AZ 85721, USA\\
$^{4}$ Department of Astronomy, University of Geneva, Ch. Maillettes 51, 1290 Versoix, Switzerland\\
$^{5}$ Department of Physics and Astronomy, University of Oklahoma, Norman OK 73019\\
$^{6}$ NuGrid collaboration, \url{https://nugrid.github.io}
}

\date{Accepted XXX. Received YYY; in original form ZZZ}

\pubyear{2020}

\begin{document}
\label{firstpage}
\pagerange{\pageref{firstpage}--\pageref{lastpage}}
\pagenumbering{arabic}
\maketitle

\begin{abstract}
In this work, we investigate the impact of uncertainties due to convective boundary mixing (CBM), commonly called `overshoot', namely the boundary location and the amount of mixing at the convective boundary, on stellar structure and evolution. For this we calculated two grids of stellar evolution models with the \texttt{MESA} code, each with the \textit{Ledoux} and the \textit{Schwarzschild} boundary criterion, and vary the amount of CBM. We calculate each grid with the initial masses $15$, $20$ and $25\,\Msun$. We present the stellar structure of the models during the hydrogen and helium burning phases. In the latter, we examine the impact on the nucleosynthesis. We find a broadening of the main-sequence with more CBM, which is more in agreement with observations. Furthermore during the core hydrogen burning phase there is a convergence of the convective boundary location due to CBM. The uncertainties of the intermediate convective zone remove this convergence. The behaviour of this convective zone strongly affects the surface evolution of the model, i.e. how fast it evolves red-wards. The amount of CBM impacts the size of the convective cores and the nucleosynthesis, e.g. the $^{12}$C to $^{16}$O ratio and the weak s-process. Lastly, we determine the uncertainty that the range of parameter values investigated introduce and we find differences of up to $70\%$ for the core masses and the total mass of the star.
\end{abstract}


\section{Introduction}

Convection is one of the key physical processes in stars and it has been studied for almost a century \citep[e.g.][]{Prandtl1925}. Nevertheless, it is still an issue to large uncertainties, among them the treatment of convective boundaries, and the stellar community only starts to understand these details of convection \citep[e.g.][]{Arnett2019a, Arnett2018b}.\\
Convection is the major contributor of turbulent energy transport, it shapes the interior of stars  and strongly influences their evolution \citep[e.g.][]{Kippenhahn1994, Woosley2002, Maeder2009}. Furthermore, it mixes the composition. This brings freshly synthesised material to the outer layers and ingests additional fuel into active burning regions. Consequently, the core of the star is more massive and the burning stage lasts longer. Moreover, it may make possible new nuclear channels due to the availability of different seed nuclei. At the convective boundaries, new material from the convectively stable region is turbulently entrained into the convective zone, which can also lead to a growth of the convective region \citep[e.g.][]{Cristini2017}. In low-mass stars, convective boundary mixing during the asymptotic giant branch phase is crucial for the creation of the $^{13}$C pocket, the s-process site in low- and intermediate mass stars \citep{Herwig2000}. In massive stars, several recent studies have shown the sensitivity of the pre-supernova structure and their explosion likelihood to the details of their complex convective history \citep{Ugliano2012, Sukhbold2014, Mueller2016, Sukhbold2016, Ertl2016, Sukhbold2018, Chieffi2019}. Yet, despite the importance of convection, these processes are still not well understood and 1-dimensional (1D) stellar evolution codes use some parametrised theory, often the mixing-length theory \citep[MLT;][]{Vitense1953, Boehm-Vitense1958}. The well-known missing shortcomings of the MLT at the convective boundary are patched together by some parametrised physics, commonly referred to as `overshooting' or `semiconvection'.\\
Convection clearly is a 3-dimensional (3D) process and with increasing computing power, simulations in 3D became possible, allowing to properly study convection \citep[e.g.][to name a few]{Herwig2006, Meakin2007, Magic2013, Woodward2015, Cristini2017, Freytag2017, Jones2017}. However, it is currently not possible to simulate the evolution of a star in 3D because the convective timescale is several orders of magnitudes shorter than the stellar lifetimes. Therefore, in order to study stellar evolution, 1D stellar evolution codes are necessary. Furthermore, 1D stellar evolution models are used as input for the 3D simulations. Therefore, a lower uncertainty in the input model would decrease the variation in the 3D simulations.\\
One longstanding conundrum in all 1D stellar evolution codes is the treatment of convective boundaries. In this work, we investigate the relative importance of the modelling uncertainties linked with convective boundary mixing (CBM) and their impact. In particular, we focus on the location of the convective boundary (`\textit{Schwarzschild} versus \textit{Ledoux} criterion') and the amount of CBM. We call the mixing beyond the convective boundary CBM to keep open the physical processes responsible for the mixing rather than calling it `overshoot', which is a specific physical process (vertical motion driven by buoyancy). Semiconvection, however, is mentioned separately because this mixing process only occurs in the models applying the Ledoux boundary criterion (see Section \ref{Theory} and \ref{convectiveUncertainties}).\\ This work focusses on the early stellar stages, starting at the zero-age main-sequence (ZAMS) up to core helium depletion. The goal of this study is (i) to highlight, which aspects of the convective boundary physics lead to the largest uncertainties in the model prediction as well as (ii) which observational test and 3D hydrodynamic simulation may help constrain convective modelling in 1D stellar evolution models. We do not use any `new' physics nor do we claim to use `right' physics. We simply use the choices that are frequently found in the literature. This study therefore helps to estimate the uncertainty of model predictions found in the literature.\\
The paper is structured as follows. In Section \ref{Theory}, we shortly review the treatment of convection in 1D stellar evolution codes. In Section \ref{physicalIngredients}, we outline the input physics and numerics used for the simulations. Furthermore, we describe the uncertainties we investigate. In Sections \ref{HydrogenBurning} to \ref{BSG_vs_RSG}, we present the impact of the variations on the stellar models and their evolution. Finally, in Section \ref{discussion}, we discuss our results and compare some quantities to the literature and in Section \ref{conclusions}, we give our conclusions.

\section{Convection in 1D Stellar Evolution}\label{Theory}

Standard 1D stellar evolution models are calculated in spherical symmetry and the convective energy transport is approximated with the MLT \citep{Vitense1953, Boehm-Vitense1958} or a theory based thereupon \citep[e.g.][]{Unno1967, Arnett1969, Spiegel1971, Canuto1991, Canuto2011}. This, however, neglects several important facts of turbulent convection, such as the convective boundary \citep{Renzini1987}.\\
The MLT is applied to the convectively unstable region to determine the convective energy flux, convective velocity and temperature gradient. However, the location of the convective boundary, i.e. the location where the buoyancy changes sign, is not part of the MLT and an additional criterion has to be used to determine this location. The two criteria mostly used in the literature are the \textit{Ledoux} and \textit{Schwarzschild} criteria.\\
The \textit{Ledoux} criterion for stability \citep[e.g. used by][]{Heger2000, Heger2005, Brott2011, Limongi2018a}, which is based on linear perturbation arguments (and large fluctuations), is formulated as $\nabla_{\rm{rad}} < \nabla_{\rm{ad}} + \frac{\varphi}{\delta} \nabla_\mu \equiv \nabla_{\rm{L}}$, with the two thermodynamic variables $\varphi \equiv (\partial \ln \rho / \partial \ln \mu)$ (at constant pressure and temperature) and $\delta \equiv (\partial \ln \rho / \partial \ln T)$ (at constant pressure and chemical composition). The $\nabla$s are temperature gradients, where $\nabla_{\rm{rad}}$ is the temperature gradient of the environment in case of pure radiative energy transport, $\nabla_{\rm{ad}}$ the temperature gradient of the MLT "bubble" as it moves and $\nabla_\mu \equiv (\partial \ln \mu / \partial \ln P)$ is the
chemical composition gradient of the surrounding.\\
The \textit{Schwarzschild} criterion for stability \citep[e.g. used by][]{Ekstroem2012, Pignatari2016, Ritter2018} is formulated as $\nabla_{\rm{rad}} < \nabla_{\rm{ad}}$.\\
The \textit{Ledoux} and \textit{Schwarzschild} criterion are the same and a difference only arises in regions with a chemical composition gradient, where the $\mu$-gradient is not equal zero. Regions that are unstable according to the \textit{Schwarzschild} criterion but stable according to the \textit{Ledoux} criterion undergo semiconvective mixing \citep[e.g.][]{Kato1966}.\\
Some studies, e.g. \citet{Georgy2014}, indicate that the \textit{Ledoux} criterion is preferred by 1D stellar models but their solution is not unique. In a purely linear theory, as the two stability criteria are, it is correct to use the \textit{Ledoux} criterion in order to take care of possible damping due to chemical composition gradients.\\
Convection, however, is a 3D process which drives intermittency and fluctuations, which are non-linear. Therefore, the boundary is not a stiff location but it bends and stretches \citep{Cristini2017}. Consequently, the chemical composition gradient is erased and the boundary becomes more similar to the \textit{Schwarzschild} boundary \citep{Meakin2007, Arnett2019a}. This is an initial value problem; the convective boundary location of the growing instability starts as the \textit{Ledoux} location and moves to the \textit{Schwarzschild} location on a finite timescale. Thus, it is not sure which criterion has to be used for convective regions that only exist on a short timescale. This behaviour needs some future 3D simulations to verify and to test the transition speed, i.e. how long the convective boundary stays at the \textit{Ledoux} location.\\
Schematically, the 3D convective boundary consists of different regions \citep[e.g.][]{Arnett2015}:
\begin{compactenum}[(a)]
\item Tubulent convective region; here the superadiabatic excess is positive and the material is unstable due to buoyant driving.
\item Turning region; as a consequence of a pressure excess and the buoyancy force changing sign, the turbulent flow turns around. This region is well mixed.
\item Shear region; the horizontal velocity dominates and the radial velocity is going to zero. The horizontal flow may create Kelvin-Helmholtz instabilities which entrain material from the stable region into the convective region.
\item Stable ("radiative") region; gravity waves are generated (as a result of the convective flow joining the stable region) but no mixing otherwise.
\end{compactenum}
These layers are not stiff but are subject to fluctuations and therefore are dynamic \citep{Cristini2017}. Recent 3D simulations of turbulent convection \citep[e.g.][]{Meakin2007, Woodward2015, Cristini2017, Jones2017} show that there is a turbulent convective, a turning and a shear region.\\
In a 1D prescription of convection using the MLT the velocity at the boundary drops to zero from a finite value, as if rammed into a solid wall. This problem arises because the MLT only considers regions (a) and (d) \citep{Renzini1987}. In order to account for CBM, i.e. regions (b) and (c), an additional theory has to be used, patching together the mixing after the convective boundary.\\
In 1D only the radial velocity is considered. Consequently, convection is often thought of as a radial up-down movement. Since the radial velocity in a 1D simulation using the MLT is not zero at the convective boundary (due to the missing turning region), the fluid overshoots the convective boundary, which is a dynamical consequence of the Newtonian laws \citep{Canuto1998}. Hence, the idea of `overshoot' was born, an attempt to locally account for CBM \citep[e.g.][]{Shaviv1973, Maeder1975}.\\
Observations indicate that CBM exists \citep[see e.g. discussion in][]{Zahn1991}. For example, CBM in stellar models is necessary in order to reproduce observations, such as the main-sequence (MS) width \citep[e.g.][the first and latter for low- and intermediate mass stars]{Maeder1975, Bertelli1984, Ekstroem2012} and asteroseimic observations \citep[e.g.][]{Straka2005, Meynet2009, Moravveji2015, Moravveji2016, Arnett2017}\\ 
Currently, there exist different implementations in 1D stellar evolution codes to account for the mixing after the 1D convective boundary. The most commonly used implementations are (i) the convective penetration \citep[e.g.][]{Zahn1991} or penetrative `overshoot' (commonly referred to as `step-overshoot') and (ii) the exponentially decaying diffusive boundary mixing \citep{Herwig1997}. The first prescription extends the fully mixed region after the convective boundary by a fraction of the pressure scale height. The second prescription is based on hydrodynamic simulations by \citet{Freytag1996} and it applies a diffusive mixing with an exponentially decreasing efficiency after the convective boundary, which inspired by the exponentially decaying velocity field seen in multi-dimensional simulations.\\
The two aforementioned CBM prescriptions may help in reproducing some observations but they do not reproduce the average shape of the complex convective boundary structure seen in 3D simulations \citep[e.g.][]{Cristini2017, Jones2017}. Furthermore, these local `add-ons' to the MLT depend on some sort of parametrisation, which result in different amount of mixing.\\
CBM still is an open question. Several new prescriptions are being developed. For example, \citet{Meakin2007} suggest a turbulent entrainment law at the convective boundary, based on 3D hydrodynamic simulations. Recently, \citet{Pratt2017} proposed a diffusion coefficient, based on a Gumbel distribution of the penetration propability in 2D hydrodynamic simulations. Some authors also combine several consisting prescriptions in order to mimic the convective boundary with the turning and the shear regions seen in 3D simulations. \citet{Michielsen2019}, for example, combine the penetrative `overshoot' and the exponentially decreasing diffusive prescription. We do not use the latter because it introduces even more free parameters.

\section{Physical Ingredients}\label{physicalIngredients}

In order to study the impact of convective boundary uncertainties in massive star models, we computed a set of non-rotating stellar models at solar metallicity with three initial masses ($15, 20, 25\,\Msun$). Our simulations were computed using the \texttt{MESA} software instrument for stellar evolution \citep{Paxton2011, Paxton2013, Paxton2015, Paxton2018}, revision 10108.\\
The radiative opacities were calculated using the tables of \citet{Asplund2009} and if $\log\,T_{\rm{eff}} \leq 3.8\,$K the opacity tables from \citet{Ferguson2005} with photospheric metals from \citet{Asplund2009} were used.\\
In order to account for the thermonuclear reactions we used a network consisting of 206 isotopes from hydrogen up the iron group. This network calls all possible reactions and their rates for its isotopes, including the weak reactions. Therefore it is suitable to calculate the energy generation for all the main burning stages during stellar evolution. Furthermore, the 206 isotope networks allows to calculate the stellar evolution up to core-collapse, it contains most of the reactions that affect the structure, such as $\alpha$-captures on $^{14}$N and $^{22}$Ne during helium burning (see discussion in Section \ref{heliumBurning}), and is able to properly calculate the neutronisation of the matter in the core, given by the electron fraction $Y_e$. Lastly, \citet{Farmer2016} showed that key quantities of the stellar models converge at the 10\% level when using an isotope network of at least $\sim 127$ isotopes. The stellar models with no CBM ($\fcbm = 0.0$, Eq.~(\ref{expD})) were calculated with a truncated network because we only use them for comparison reasons. The truncated network consists of all the elements up to aluminium in \texttt{mesa\_206.net} and additionally silicon $27,28,29$. It is therefore suitable to calculate all the necessary reactions during the hydrogen and helium burning phases. The truncated network introduces no difference during the core hydrogen and helium burning stages. The reaction rates are taken from the \texttt{JINA REACLIB} \citep{Cyburt2010}.\\
The initial metal elemental abundances were taken from \citet{Asplund2009} with some elements (He, C, N, O, Ne, Mg, Al, Si, S, Ar, Fe) updated based on \citet{Nieva2012} and \citet{Przybilla2013}.\\
We accounted for mass loss by stellar winds with \texttt{MESA}'s \texttt{Dutch} mass loss scheme. This includes several prescriptions; for O-stars the mass loss rates from \citet{Vink2000, Vink2001} are used. If the star enters the Wolf-Rayet stage, i.e. when the surface hydrogen mass fraction drops below $0.4$, the mass loss rate switches to the scheme from \citet{Nugis2000}. If $T_{\rm{eff}} < 10^4\,$K, the empirical mass loss rate from \citet{deJager1988} was used. All the mass loss rates were scaled with a factor of 0.85. This reduction factor was introduced by \citet[][see Section 2.2 for details]{Maeder2001} or empirical mass loss rates. While this reduction factor is not necessary for theoretical mass loss rates such as \citet{Vink2000, Vink2001}, we used it for all phases to have mass loss rates similar to published GENEC models \citep[GENEC applies the factor 0.85 during the MS, e.g.][]{Ekstroem2012} and MESA models \citep[e.g.][apply a factor of 0.8]{Farmer2016, Ritter2018}.\\
Some of the models generate enough luminosity so that their radiation pressure dominated envelope experience gas pressure and density inversion \citep[e.g.][]{Joss1973}. These models become numerically unstable and the timesteps become prohibitively short (of the order of hours). In order to keep the numerics stable and the timsteps at a reasonable limit we use \texttt{MESA}'s \texttt{MLT++} \citep[][Section 7.2]{Paxton2013} in all models that apply the largest amount of CBM and in the $20$ and $25\,\Msun$ models also with the second largest amount (see below). The treatment of \texttt{MLT++} allows the calculation of these models to the end of core helium burning with reasonable timesteps. Tests of the \texttt{MLT++} formalism in $15\,\Msun$ models do not show any significant differences in the structure and evolution but see discussion in Section \ref{discussion}.\\
The MESA models assume hydrostatic equilibrium and apply the MLT variation of \citet{Henyey1965}. The mixing length was set to $\ell_{\rm{MLT}} = 1.6\, H_P$, where $H_P$ is the pressure scale height. This is the same value used by \citep{Ekstroem2012}. Furthermore, for strongly stratified convection \citet{Arnett2018b} find an asymptotic limit for the dissipation length of a turbulent flow, which they identify with $\ell_{\rm{MLT}} \sim H_\rho \sim 5/3\, H_P$, which is close to $1.6\, H_P$.  The mixing of the nuclear species in \texttt{MESA} is assumed to be a diffusive process. The diffusion coefficient in the convective region is calculated by $D = \frac{1}{3} \ell_{\rm{MLT}} v_{\rm{MLT}}$, where $v_{\rm{MLT}}$ is the velocity determined by the MLT.\\
We use the same resolution, at which our models seem to converge, in all calculations except the $15\,\Msun$ models with no CBM. In these models, we needed to increase the resolution in order to properly resolve the the boundary of the convective zones. The details can be found in the inlists\footnote{\href{http://doi.org/10.5281/zenodo.3871897}{http://doi.org/10.5281/zenodo.3871897}}.

\subsection{Convective Boundary Mixing Uncertainties}\label{convectiveUncertainties}

In this study, we investigate two uncertainties due to CBM, (i) the determination of the convective boundary location and (ii) different amounts of extra mixing after the convective boundary.\\
As discussed in Section \ref{Theory}, the determination of the convective boundary is not included in the MLT and either the \textit{Ledoux} or the \textit{Schwarzschild} criterion has to be used. We calculated every model twice, once applying the \textit{Ledoux} and once the \textit{Schwarzschild} criterion to address this uncertainty.\\
An investigation of the second point is a much more extensive task because CBM is poorly understood, hence connected with several uncertain aspects. The uncertainties arise from (a) the poor knowledge of the convective boundary and the breakdown into 1D, thus, how to describe and implement the physics in 1D, (b) the parametrisation of the CBM prescriptions and (c) the different implementations of the same theory in the various stellar evolution codes \citep[see e.g. discussions in][]{Jones2015, Stancliffe2016}. In this work, we limit ourselves to one CBM prescription and investigate the impact of different choices of the free parameters within this setting.\\
\citet{Moravveji2015} tested the penetrative `overshoot' and exponentially decaying `overshoot' against asteroseismic observation. They found a better fit with the exponentially decaying `overshoot' prescription. Furthermore, \citet{Arnett2017} show that the asteroseismic models from \citet{Moravveji2015, Moravveji2016} with the exponentially decaying `overshoot' prescription create a chemical composition profile similar to the profile in 3D hydrodynamic simulations. On the contrary, the penetrative `overshoot' creates a step in the chemical composition profile, which is only seen in 1D stellar models. The different chemical composition profile results in a different local structure and finally affects, for example, the boundary criterion. Therefore, we chose to use the exponentially decaying `overshoot' formalism in this work. We note that the free parameters in the two prescriptions can be mapped with a mapping factor between $10-15$ \citep{Herwig1997, Noels2010, Moravveji2016, Claret2017}.\\
The exponentially decaying CBM prescription is based on hydrodynamic simulations by \citet{Freytag1996}. Since we simulated the interior, where the instabilities at the convective boundary behave different than in the surface convection simulations from \citet{Freytag1996}, we refer to the resulting mixing after the convective boundary as CBM. This includes an ensemble of different physical processes which might cause mixing across the convective boundary and is not only limited to an `overshooting' of the convective flow at the boundary. Even if the convective flow is simulated as a radial up-down movement in 1D stellar evolution it is still necessary to think of convection as a 3D process.\\
The diffusion coefficient of the exponentially decaying CBM is calculated as \citep{Herwig1997}

\begin{equation}
	D_{\rm{CBM}} = D_0(f_0) \cdot \exp\left( \frac{-2 z}{\fcbm \cdot H^{\rm{CB}}_{P}} \right).\label{expD}
\end{equation}

The diffusion coefficient is a function of distance $z = r - r_0(f_0)$  from a point close to the edge of the convective boundary. $\fcbm$ is a free parameter which expresses the distance of the extra mixing as a fraction of the pressure scale height at the convective boundary, $H^{\rm{CB}}_{\rm{P}}$. $D_0(f_0)$ is the diffusion coefficient from the MLT, taken at the location $r_0(f_0) = r_{\rm{CB}} - f_0 \cdot H_P$ inside the convective zone, where $r_{\rm{CB}}$ is the location of the convective boundary determined by the boundary criterion and $f_0$ is an additional free parameter\footnote{We note that the implementation of the penetrative `overshoot' in the \texttt{MESA} code depends on a similar second parameter.}. This is done because the diffusion coefficient from MLT drops sharply towards zero at the convective boundary. The new diffusion coefficient is then applied starting at $r(f_0)$, thus inside the convective zone.\\
In the CBM zone, the temperature gradient is set equal to the radiative one. The chemical composition, on the other hand, is mixed using the diffusion coefficient determined by Eq.\,(\ref{expD}). The diffusive mixing after the convective boundary is cut-off at a certain value, which we chose to be $D_{\rm cut} = 10^2\,$cm$^2\,$s$^{-1}$, in order to avoid the long exponential tail. This treatment of CBM is applied to all boundaries of all convective zones.\\
$D_0$ has to be taken `close' to the edge of the convective boundary \citep{Herwig2000},  which is equivalent to a small $f_0$ parameter. It is often not discussed how `close' and only the $\fcbm$ parameter is mentioned, despite the importance of $f_0$. Changing the $f_0$ parameter in Eq.~(\ref{expD}) from $0.02$ to $0.002$, gives a different location (i) where $D_0$ is taken from and (ii) where to begin the exponential decrease of the diffusion coefficient. The impact of the first point is negligible since the MLT predicts an approximately constant diffusion coefficient. The second point, however, is not negligible. The fact that the diffusion coefficient begins to decrease deeper in the convective region and is cut off after it drops below a certain value means that the mixing efficiency inside the convective zone recedes and there is less and weaker mixing after the convective boundary.\\
The CBM model prescription is used regardless of possible chemical composition gradients at the convective boundary. Those might affect the amount of CBM but will not prevent it entirely \citep{Canuto1998}.\\
\citet{Herwig2000} find that $\fcbm=0.016$ is needed for convective core hydrogen burning in intermediate mass stars to reproduce the MS width. \citet{Claret2017, Claret2018} do a semi-empirical mass calibration of $\fcbm$ and find a dependence of $\fcbm$ on the stellar mass, with a strong increase of $\fcbm$ up to about $2\,\Msun$ where it levels off at a value of $\fcbm \sim 0.0164 - 0.0181$. \citet{Denissenkov2019}, on the other hand, scale the $\fcbm$ with the driving luminosity, $\fcbm \propto \rm{L}^{1/3}$.\\
CBM is often constrained with observations using the penetrative `overshoot' in the stellar evolution calculations, where the fully mixed region is extended by a fraction of the pressure scale height, $\alpha_{\rm ov} \cdot H_P$. In this case, $\alpha_{\rm ov}$ is the free parameter which is constrained. \citet{Ekstroem2012} fit their amount of CBM to low mass stars with $\alpha_{\rm ov} = 0.1$. \citet{Brott2011} constrain CBM with the observed drop of the rotation rates for stars with a surface gravity of $\log\,g < 3.2$ and find $\alpha_{\rm ov} = 0.335$ for a $16\,\Msun$ star. Recently, \citet{Schootemeijer2019} compare a grid of stellar models with varying amounts of internal mixing to observations of massive stars in the Small Magellanic Cloud and conclude that $0.22 \lesssim \alpha_{\rm ov} \lesssim 0.33$ is needed to match observations of blue to red supergiants. \citet{Costa2019} reanalyse the sample of \citet{Claret2017, Claret2018} and find a wide distribution of $0.3-0.4 < \alpha_{\rm ov} < 0.8$ for masses $M > 1.9\,\Msun$ in non-rotating models. When they include rotation, the models agree with the observed data when $\alpha_{\rm ov} = 0.4$. \citet{Higgins2019}, on the other hand, constrain massive star evolution with a galactic binary system. They need $\alpha_{\rm ov} = 0.5$ (and rotation) in order to reproduce the system. These parameters map to $\fcbm = (0.0335, 0.04, 0.05)$ by using a mapping factor of $\sim 10$. Considering the uncertainty in the 1D mixing process, this approximation is acceptable.\\
\citet{Castro2014} suggest, based on observational spectroscopic Herzsprung-Russell diagram (HRD) of Galactic massive stars, that the amount of CBM increases with initial mass in order to fit their empirical terminal-age MS. Other studies \citep[e.g.][]{Vink2010, McEvoy2015}, however, do not find a clear boundary corresponding to the terminal-age MS. Nevertheless, \citet{McEvoy2015} find hints for a broader MS width than adopted in the literature. Grin et al. (in prep.) find that the best fit to the empirical MS-width in the initial mass range $8 - 22\,\Msun$ with an increasing $\alpha_{\rm ov}$ from $0.2$ to $0.5$ for non-rotating models. For models with initial masses of $15$ and $20\,\Msun$ they use $\alpha_{\rm ov} \sim 0.5$.\\
The values commonly used in theoretical `\textit{state-of-the-art}' evolution calculations of massive stars range from $\fcbm = 0.004$ \citep[e.g.][]{Farmer2016, Fields2018} up to $\fcbm = 0.022$ \citep[e.g.][]{Jones2015} or $\fcbm = 0.025$ \citep[e.g.][]{Sukhbold2014}, with intermediate values around $\fcbm = 0.014 - 0.016$  \citep[e.g.][]{Choi2016, Pignatari2016, Ritter2018}. These values are lower than the values for massive stars constrained by observations and the difference will influence the structure and evolution of the star.\\
In order to cover the range of $\fcbm$ adopted in the literature and the constraints from observations we used the values $(0.004, 0.01, 0.022, 0.035, 0.05)$. Moreover, for comparison, we also calculated all the models with an initial mass of $15\,\Msun$ with no CBM. Additionally, we test two values for $f_0$, $0.002$ and $0.02$, in the $15\,\Msun$ models with $\fcbm \leq 0.022$. We limit ourselves to the small values of $\fcbm$ because the relative importance of $f_0$ becomes negligible in models with large amounts of CBM (see e.g. Table \ref{properties_hedep}). The $20$ and $25\,\Msun$ models are only simulated with $f_0 = 0.002$.\\
Regions which are unstable according to the \textit{Schwarzschild} criterion but stable according to the \textit{Ledoux} criterion undergo slow semiconvective mixing. We used the semiconvective prescription from \citet{Langer1983} who formulate the semiconvective mixing as a diffusive process. The diffusion coefficient is calculated as

\begin{equation}
	D_{\rm{sc}} = \alpha_{\rm{sc}} \frac{K}{6C_P\rho} \frac{\nabla - \nabla_{\rm{ad}}}{\nabla_{\rm{L}} - \nabla}. \label{semiconvectionFromulae}
\end{equation}

$K$ is the radiative conductivity and $C_p$ the heat capacity at constant pressure. The semiconvective diffusion coefficient is further scaled by the semiconvective efficiency parameter $\alphasc$.\\
The amount of semiconvective mixing, if it occurs, is still an unsolved problem \citep[e.g.][and references therein]{Langer2012}, hence $\alphasc$ is uncertain. \citet{Langer1985} estimate the semiconvective efficiency to be of the order of $0.1$. The values used in the literature vary greatly, ranging from small values of $\alphasc = 0.01-0.02$ \citep[e.g.][]{Farmer2016, Limongi2018a} up to $1.0$ \citep[e.g.][]{Brott2011}, with intermediate values of $\sim 0.1$ \citep[e.g.][]{Sukhbold2014, Choi2016}. \citet{Schootemeijer2019} explore in their calculations a large range of $\alphasc = 0.01 - 300$ and conclude that $\alphasc > 1.0$ is needed to reproduce the BSG to RSG ratio in the Small Magellanic cloud.\\
We used two values for $\alphasc$, $0.4$ (fast semiconvection) and $0.004$ (slow semiconvection) in our $15\,\Msun$ models. The $20$ and $25\,\Msun$ models where only calculated with $\alphasc = 0.4$ because the relative importance of semiconvection decreases with increasing amount of mixing at the convective boundary (see Sections \ref{HydrogenBurning}, \ref{ICZ} and \ref{heliumBurning}). Therefore, the two values of $\alphasc$ would predict a similar outcome. Similarly, \citet{Schootemeijer2019} find that in their massive star models of the Small Magellanic Cloud semiconvection rarely develops for large amount of CBM  and only plays a role after the MS. Moreover, \citet{Langer1985} show that while semiconvection can occur prominently during the MS evolution in their massive star models the evolution during this phase is nearly independent of the choice of $\alphasc$.

\section{Core Hydrogen Burning} \label{HydrogenBurning}

In the core hydrogen burning phase, hydrogen is fused into helium. This increases the mean molecular weight $\mu$ and decreases the opacity $\kappa$. The first leads to an increase in luminosity, because $L \propto \mu^4$ \citep[e.g.][]{Kippenhahn1994}, hence, a reduction of the pressure onto the core. The decrease of the opacity and pressure dominate over the increase of the core luminosity in a massive star. Therefore, since $\nabla_{\rm{rad}} \propto \kappa \ell_{\rm{rad}} P$ \citep[e.g.][]{Kippenhahn1994}, the radiative temperature gradient decreases. On the other hand, the adiabatic temperature gradient, $\nabla_{\rm{ad}}$ remains roughly constant in the interior of the star. This constantly stabilises the material at the convective boundary against convection according to the stability criterion, and the mass of the convective hydrogen in a massive star decreases during the MS lifetime. A consequence of the decreasing convective core is a decreasing mean molecular weight above the convective core. The resulting $\mu$-gradient creates the difference between the two boundary criteria.\\
\begin{figure}
	\includegraphics[width=\columnwidth]{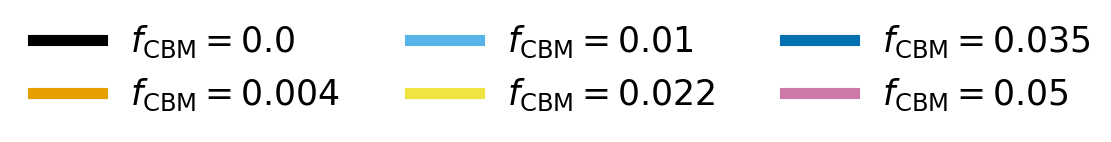}
	\includegraphics[width=\columnwidth]{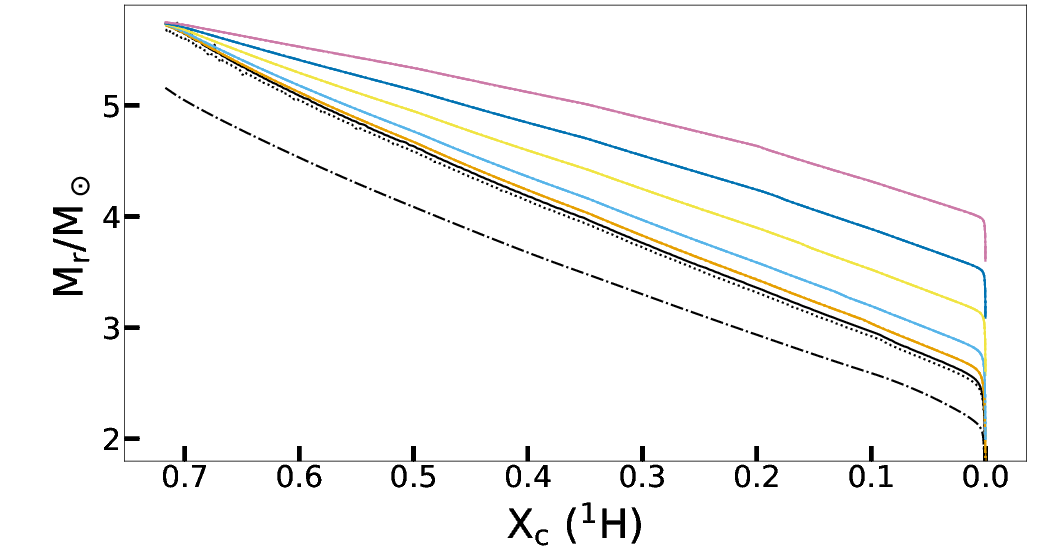}
	\caption{The location of the convective hydrogen core boundary, determined by either the \textit{Ledoux} or \textit{Schwarzschild} criterion, as a function of the central hydrogen mass fraction. All tracks are $15 M_\odot$ models with $f_0 = 0.002$. The solid lines indicate Schwarzschild models and the other lines are Ledoux models with either $\alphasc = 0.4$ (dotted line) or $\alphasc = 0.004$ (dash-dotted line). The color scheme shows the different choices of $\fcbm$. The inset window presents the evolution of the convective boundary location right before it reaches the ZAMS.}
	\label{boundaryLocation_hydrogenCore}
\end{figure}
Fig.~(\ref{boundaryLocation_hydrogenCore}) shows the location of the convective boundary in stellar evolution models with an initial mass of $15\,\Msun$, either given by the \textit{Ledoux} or the \textit{Schwarzschild} boundary criterion, for various amounts of extra mixing and $f_0 = 0.002$. The location presented in Fig.~(\ref{boundaryLocation_hydrogenCore}) is the pure \textit{Ledoux} or \textit{Schwarzschild} boundary excluding the CBM region. It is apparent, that the location of the convective boundary is further out in mass coordinates with more CBM. This is a consequence of the larger mixed region after the convective boundary. The inset window in Fig.~(\ref{boundaryLocation_hydrogenCore}) presents a zoom on the final growth of the convective hydrogen core before the ZAMS. It shows that all the $15\,\Msun$ models, except the Ledoux model with no CBM and slow semiconvection, have a nearly equal convective hydrogen core size at the ZAMS. Therefore, the differences arising during the MS evolution are due to the larger $\fcbm$ values. More CBM increases the overall size of the convective zone, ingesting more fuel into the burning zone in the centre. This creates a higher hydrogen burning luminosity. Consequently the decrease in the radiative temperature gradient is relatively slower, which results in a larger convective hydrogen core (see also Table \ref{properties_hedep}).\\
\begin{figure*}
	\includegraphics[width=\linewidth]{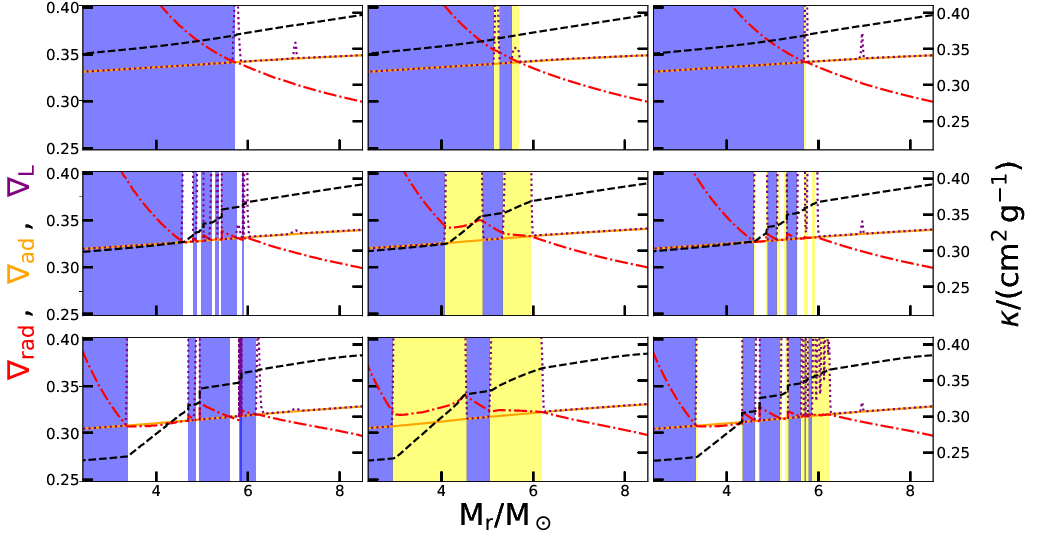}
	\caption{Profiles of the temperature gradients at the boundary of the convective core as a function of Lagrangian mass coordinates. Shown are the $15 M_\odot$ Schwarzschild (left column) and the Ledoux models with $\alpha_{\rm{sc}} = 0.004$ (middle column) and $\alpha_{\rm{sc}} = 0.4$ (right column), all with no CBM. The top row is at the ZAMS (X$_c$($^1$H) $= 0.717$), the middle row at X$_c$($^1$H) $= 0.5$ and the bottom row at X$_c$($^1$H) = $0.2$. Convective regions are indicated by blue shading, whereas yellow shows semiconvective regions. Additionally, the opacity is plotted as a function of mass coordinate (black dashed line). The $\nabla_{\rm{L}}$ in the Schwarzschild model is only included for comparison and not used in the calculation.}
	\label{temperatureGradients_15M_f0p0}
\end{figure*}
The models with no CBM (black lines in Fig.~(\ref{boundaryLocation_hydrogenCore})) predict different locations of the convective boundary by either using the \textit{Ledoux} or the \textit{Schwarzschild} criterion. In the Schwarzschild model the chemical composition gradient is ignored. Therefore, the convective core can grow freely during the pre-MS evolution. The Ledoux models estimate a different boundary location depending on the semiconvective efficiency. The Ledoux model with inefficient semiconvection ($\alphasc = 0.004$) shows a convective core which is smaller. This is because during the pre-MS, where the convective core grows, a strong chemical composition gradient limits its size (see inset window in Fig.~(\ref{boundaryLocation_hydrogenCore})). A semiconvective layer develops above the convective core but semiconvection is not efficient enough to completely remove the chemical composition gradient. As a result, this model has a smaller convective core during the whole MS evolution. If semiconvection is efficient ($\alphasc = 0.4$), the $\mu$-gradient in the layer above the core is erased and the convective core can grow more. Therefore, the Ledoux model with efficient semiconvection has a convective hydrogen core size more similar to the Schwarzschild model at the ZAMS. Afterwards, during the MS evolution, the Schwarzschild model and the Ledoux model with $\alphasc = 0.4$ evolve their decreasing convective core similarly. This also is presented in Fig.~(\ref{temperatureGradients_15M_f0p0}) where the radial profile of the temperature gradients, the chemical composition gradient and the opacity are shown. The two models (left and right columns) behave similarly because in the latter there is a thin semiconvective zone (yellow) right at the convective core boundary, which constantly mixes the region above the core. The Ledoux model with $\alphasc = 0.004$ evolves through the MS with a smaller convective core and a large $\mu$-gradient above it (middle column in Fig.~(\ref{temperatureGradients_15M_f0p0})). These differences affect the helium core mass at core hydrogen depletion (Table \ref{properties_hedep}) and the luminosity during the MS (Fig.~(\ref{HRD})), which in turn impact the further evolution.\\
Fig.~(\ref{temperatureGradients_15M_f0p0}) might suggest that the Ledoux model with inefficient semiconvection (middle column) develops a chemical composition gradient within the convective zone which then is split up during the MS evolution. Careful investigation reveals that the inner (left) location is the upper limit of the convective core, whose growth is limited due to the strong, narrow peak of the $\mu$-gradient at the edge of the convective core. The convective region above the core develops during the pre-MS, which results in the convective layer after the $\mu$-gradient peak (middle column, top panel).\\
\begin{figure*}
	\includegraphics[width=\linewidth]{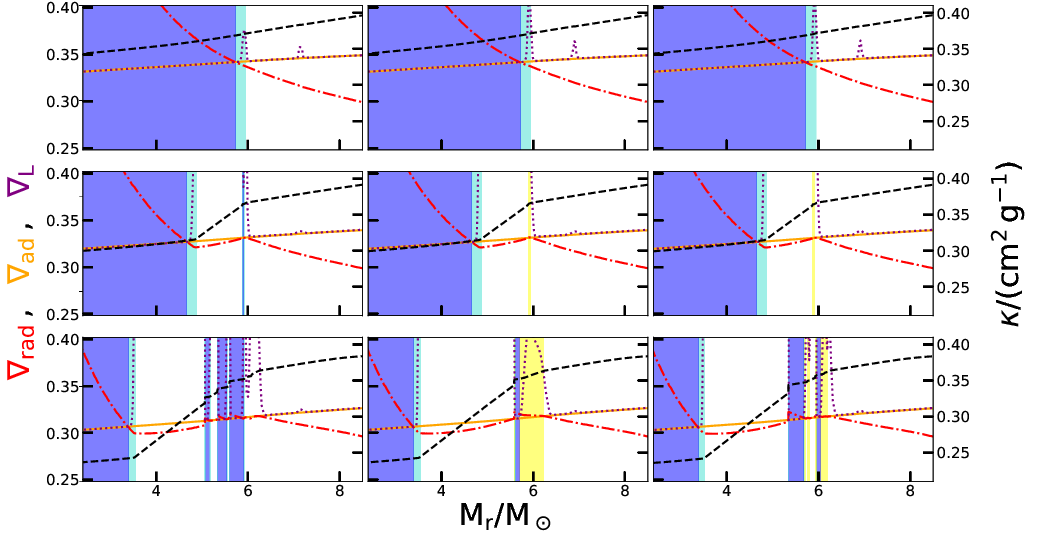}
	\caption{The same as Fig.~(\ref{temperatureGradients_15M_f0p0}) but with $\fcbm = 0.004$. The CBM region is indicated by turquoise.}
	\label{temperatureGradients_15M_f0p004}
\end{figure*}
CBM extends the region above the core that is well mixed. Consequently, changes in the chemical composition and the increase of opacity are pushed further away from the boundary location obtained from the stability criteria. Consequently, $\gradr$ decreases further after the convective boundary. Fig.~(\ref{temperatureGradients_15M_f0p004}), which shows the same stellar models as Fig.~(\ref{temperatureGradients_15M_f0p0}) but with $\fcbm = 0.004$ instead of $0.0$, illustrates this behaviour. Moreover, the chemical composition gradient at the convective boundary vanishes, $\nabla_\mu \approx 0$, and its increase is not a step-function anymore but is more sigmoid-shaped. As a result, the convective hydrogen core boundary predicted by the \textit{Ledoux} and \textit{Schwarzschild} criterion in Fig.~(\ref{boundaryLocation_hydrogenCore}) converge. Obviously, the convergence between the two stability criteria is consistent when more CBM is applied as shown in Fig.~(\ref{boundaryLocation_hydrogenCore}). The convergence of the two boundary criteria is also apparent in the HRD (Fig.~(\ref{HRD})), where the evolutionary tracks with CBM perfectly overlap during the MS. Furthermore, they predict the same helium core mass at the end of hydrogen burning (Table \ref{properties_hedep}).\\
Semiconvection only influences the convective hydrogen core size when there is no CBM in our $15\,\Msun$ models. There, a chemical composition gradient on the radiative side of the boundary limits the growth of the convective hydrogen core depending on the semiconvective efficiency (see Figs.~(\ref{boundaryLocation_hydrogenCore}) and (\ref{temperatureGradients_15M_f0p0}) with $\alphasc = 0.004$ and $0.4$). However, as discussed before, CBM removes the gradient in chemical composition on the radiative side of the convective boundary. Additionally, the radiative temperature gradient further decreases in the convective boundary region, which creates the condition $\grada - \gradr > 0$ after the CBM region (Fig.~(\ref{temperatureGradients_15M_f0p004})). For that reason there is no semiconvective region right after the convective core region when CBM is applied and the convective hydrogen core is independent of semiconvection or its efficiency.\\
The differences of the convective boundary region discussed above have an effect on the MS evolution. The larger convective cores enable more hydrogen fuel to be ingested into the central burning region. Subsequently, the helium core mass at the end of core hydrogen burning increases with more CBM (Table \ref{properties_hedep}). Furthermore, the luminosity generated by core hydrogen burning is higher and the increased radiation pressure leads to slightly larger radius of the star. The consequence is that the track in the HRD in Fig.~(\ref{HRD}) is steeper and reaches lower effective temperatures at the end of the MS (Table \ref{properties_hedep}), hence, the MS width broadens, especially for the models with large amount of CBM. Also, the increased amount of hydrogen available in the core burning region enhances the MS lifetime (Table \ref{properties_hedep}).\\
The behaviour of the convective hydrogen core and its response to CBM uncertainties found for the $15\,\Msun$ models is similar for stellar model with initial masses of $20$ and $25\,\Msun$. There is, however, a small difference in the Schwarzschild and the Ledoux model. The thin convective layers found above the convective hydrogen core (\textit{convective fingers}, see discussion further down) penetrate slightly deeper in the models with larger initial masses and sometimes touch the convective core. This transports fuel into the convective core which leads to an increase of the convective core. The timing and intensity of the `touching' is different for the Ledoux and Schwarzschild models and depends on the initial mass. Therefore, the initially converged boundary locations diverge once more and the models end up with slightly different helium core masses at the end of core hydrogen burning (Table \ref{properties_hedep}). A similar scenario is observed by e.g. \citet{Farmer2016} (their Fig.~3) for a star with an initial mass of $30\,\Msun$ and \citet{Clarkson2020}. There, however, the process is much more intense and the increase of the convective hydrogen core is larger compared to our cases. Whether such a merging scenario is realistic needs to be determined, though, with more realistic boundary physics \citep[e.g. the Richarson number instead of the \textit{Ledoux} or \textit{Schwarzschild} criterion; ][]{Turner1973} instead of simply adding the diffusion coefficients together.\\
The $20$ and $25\,\Msun$ models show a larger dispersion of minimum effective temperatures reached at the end of the MS evolution (Table \ref{properties_hedep}). This indicates that, if the $\fcbm$ value is indeed as large as recent observational calibrations, the widening of the MS width is more extreme for higher initial masses. Furthermore, the line with the terminal-age main-sequence is slightly bent towards cooler temperatures rather than to hotter temperatures with increasing initial mass as suggested by recent observations \citep{Castro2014, McEvoy2015}. Comparing the different log$_{10}\,\rm{T}_{\rm eff, min}^{\rm MS}$ values in Table \ref{properties_hedep} reveals that the MS width is nearly independent of the convective boundary criterion and the semiconvective efficiency. This is because during the MS evolution (i) there is a convergence between the two boundary criteria and (ii) the relative importance of semiconvection is massively reduced with increasing CBM (see discussion above).\\
In Section \ref{convectiveUncertainties} we mentioned the importance of another free parameter, $f_0$, in the exponentially decreasing diffusive CBM model. Changing this parameter from our default value of $0.002$ to $0.02$ decreases the amount of mixing beyond the convective core. Consequently, slightly less fuel is brought down into the burning region, resulting in a lower $\gradr$, hence, the convective boundary location  decreases faster during the MS evolution for the same $\fcbm$. This flattens the MS evolution track in the HRD and reduces the MS width. Moreover, the helium core mass at core hydrogen depletion is smaller (Table \ref{properties_hedep}). However, the differences due to the two $f_0$ values decrease with increasing $\fcbm$, because the $f_0$ is smaller relative to the $\fcbm$ parameter. Therefore, the impact of the earlier decrease of the diffusion coefficient is reduced.

\paragraph*{Convective Fingers:}\label{convectiveFingers}

In the region above the convective hydrogen core, the radiative temperature gradient has a profile close to adiabatic one ($\gradr \approx \grada$), see e.g. in Figs.~(\ref{temperatureGradients_15M_f0p0}) and (\ref{temperatureGradients_15M_f0p004}). Such a convective neutral region above the convective hydrogen core in massive stars was first predicted by \citet{Schwarzschild1958}. This is a concequence of the increasing  opacity in the region of the decreasing convective hydrogen core due to the transition to a more hydrogen-rich mixture. Accordingly, the radiative temperature gradient, $\gradr \propto \kappa$, has a flatter profile, or even slightly increases, in the region above the receding convective hydrogen core. \citet{Schwarzschild1958} propose that this zone above the convective core is slowly mixed to maintain convective neutrality. Our simulations with no CBM (Fig.~(\ref{temperatureGradients_15M_f0p0})) show a similar behaviour: At the ZAMS (top row) the temperature gradient above the convective core decreases outwards (increasing mass coordinate). During the MS evolution the convective core slowly retreats, leaving behind a composition gradient which increases the opacity. Consequently, the radiative temperature gradient above the receding convective hydrogen core is close to adiabatic. Hence, small discontinuities in the opacity profile, which create small local peaks in the radiative temperature gradient, violate the \textit{Schwarzschild} stability criterion. This results in a thin layer with mixing, that reduces the radiative temperature gradient back to the adiabatic one. At the boundary of these mixed layers, a new discontinuity in opacity is created and the process repeats itself there. This creates a finger-like structure in the region above the core \citep[e.g.][]{Langer1985}. The difference between the \textit{Ledoux} and \textit{Schwarzschild} criterion is the type of mixing in the thin layers. In the Schwarzschild models the \textit{convective fingers} are always convectively mixed. In the Ledoux models, however, the layer appears as semiconvective layer because of the strong chemical composition gradient above the receding hydrogen core. If semiconvection is not efficient a large semiconvective region develops above the convective core because the mixing is not efficient enough to completely remove the chemical composition gradient and, at the same time, the \textit{Schwarzschild}-unstable layer grows due to the receding convective hydrogen core\footnote{The convective layer in this semiconvective region seen in Fig.~(\ref{temperatureGradients_15M_f0p0}), middle column, is a relic from the pre-MS evolution, see previous discussion, and exists during the whole MS.}. If semiconvection is more efficient, it is able to remove the chemical composition gradient. It should be noted that semiconvection, as we use it, only mixes the chemical composition but ignores the thermodynamics \citep[e.g. the temperature,][]{Langer1983}. Therefore, the layer becomes convectively unstable because still $\gradr > \grada$. Thus, a similar finger-like convective-semiconvective structure as in the Schwarzschild models develops (Fig.~(\ref{temperatureGradients_15M_f0p0})).\\
CBM (i) pushes the transition from the helium-rich mixture in the convective core to the hydrogen-rich mixture in the envelope further away from the convective boundary and (ii) creates a smoother transition due to the exponential nature of the CBM. The latter creates a more continuous opacity profile, therefore a smoother $\gradr$ profile. The first point, on the other hand, causes the opacity to increase further away from the boundary. This allows the radiative temperature gradient to further decrease in the CBM region before it raises once more due to the increase of opacity. Hence, the appearance of \textit{convective fingers} is either further out (e.g. Fig.~(\ref{temperatureGradients_15M_f0p004}), middle row) or they never occur because $\gradr$ drops enough for the region above the core to stay convectively stable. Thus, the spatial area where \textit{convective fingers} occur, if any, is reduced with increasing amount of mixing at the convective boundary. For $\fcbm \gtrsim 0.01$ there are no \textit{convective fingers} in our $15\,\Msun$ models.\\
The $20$ and the $25\,\Msun$ models exhibit a similar behaviour regarding the mixing in the zone beyond the convective core as the $15\,\Msun$ models but there are some important differences. Stars with a higher initial mass generate a higher luminous output. Hence, $\gradr \propto \ell_{\rm rad}$ is much closer to convective neutrality in the radiative zone beyond the convective core. Therefore, in higher mass star models, smaller changes in the entropy immediately create a situation where the stability criteria predict convection (or semiconvection). Consequently, the \textit{convective fingers} are much more present in the simulations with the same $\fcbm$ but higher initial masses. As a result, the limit of CBM above no \textit{convective fingers} or semiconvective layers appear increases with initial mass. In the $20\,\Msun$ models we do not find \textit{convective fingers} for $\fcbm \gtrsim 0.022$ and in the $25\,\Msun$ models for $\fcbm \gtrsim 0.035$.

\section{The Intermediate Convective Zone} \label{ICZ}

\begin{figure}
	\includegraphics[width=0.49\columnwidth]{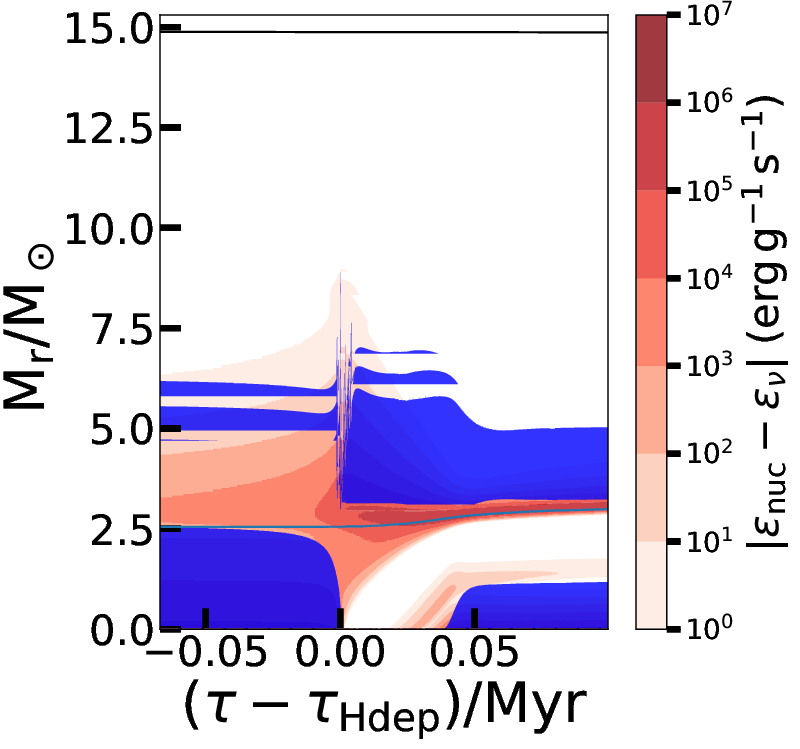}
	\hspace{0.2cm}
	\includegraphics[width=0.49\columnwidth]{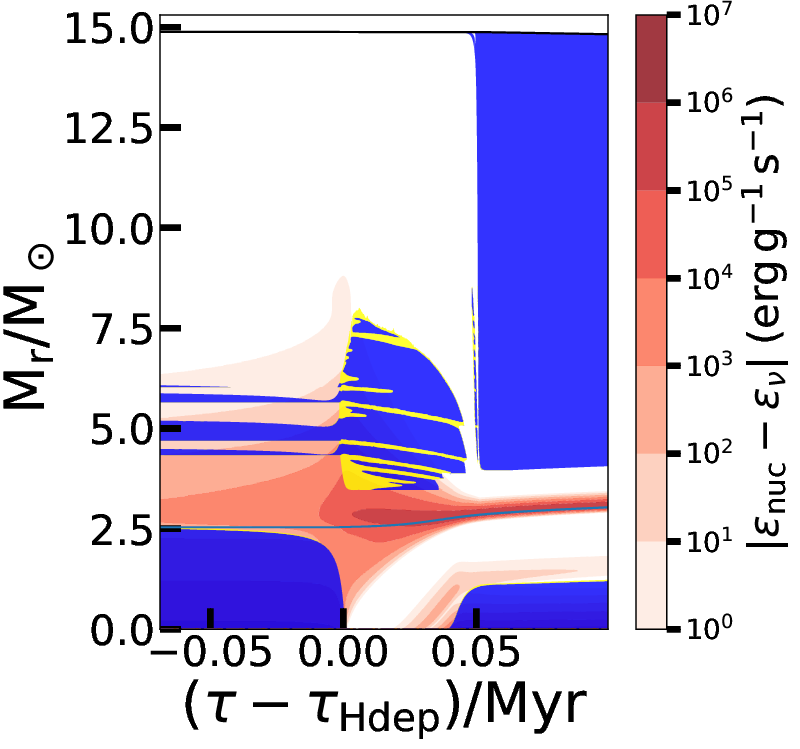}\\
	\includegraphics[width=0.49\columnwidth]{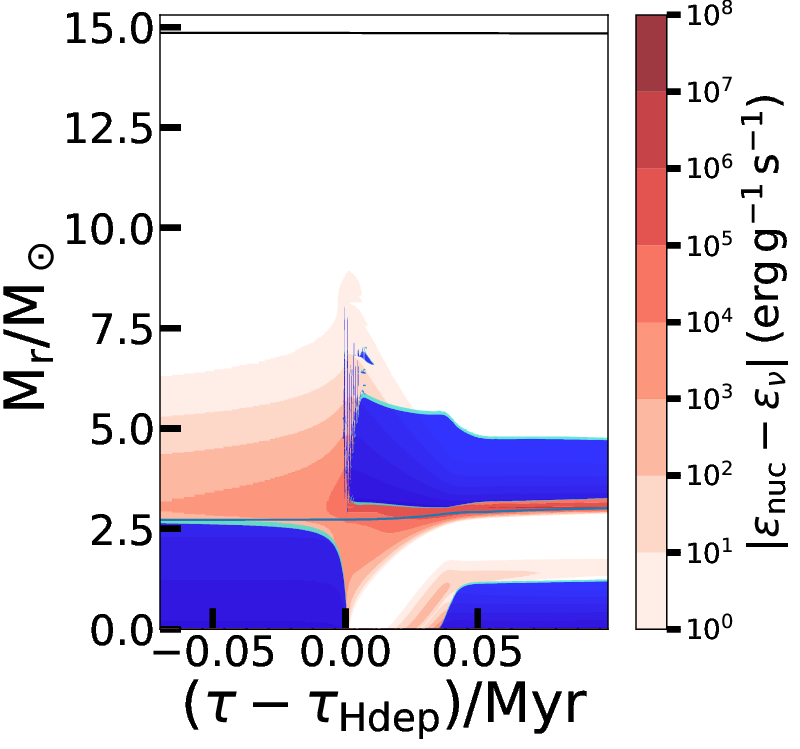}
	\hspace{0.2cm}
	\includegraphics[width=0.49\columnwidth]{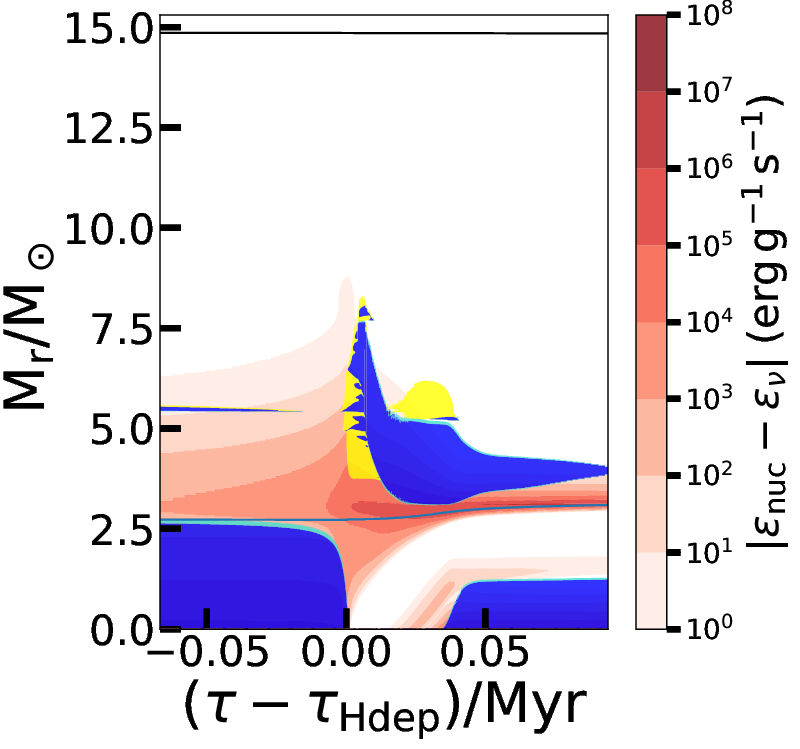}
	\includegraphics[width=0.49\columnwidth]{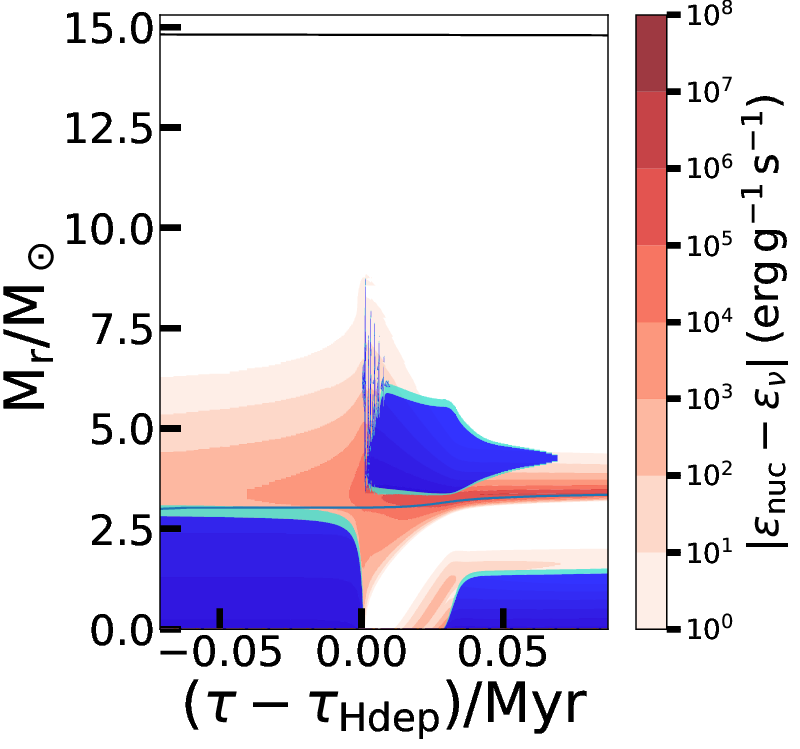}
	\hspace{0.2cm}
	\includegraphics[width=0.49\columnwidth]{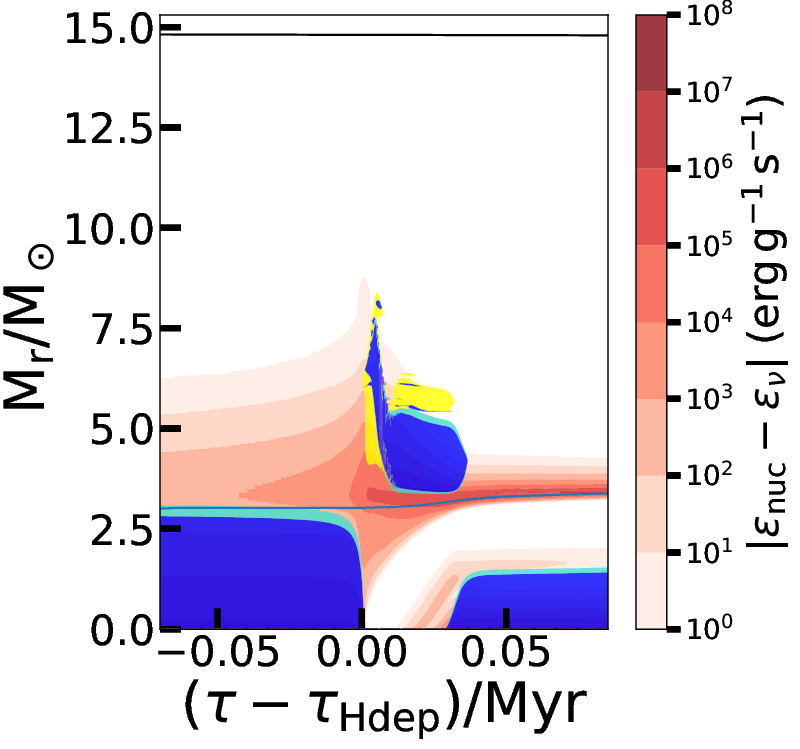}
	\includegraphics[width=0.49\columnwidth]{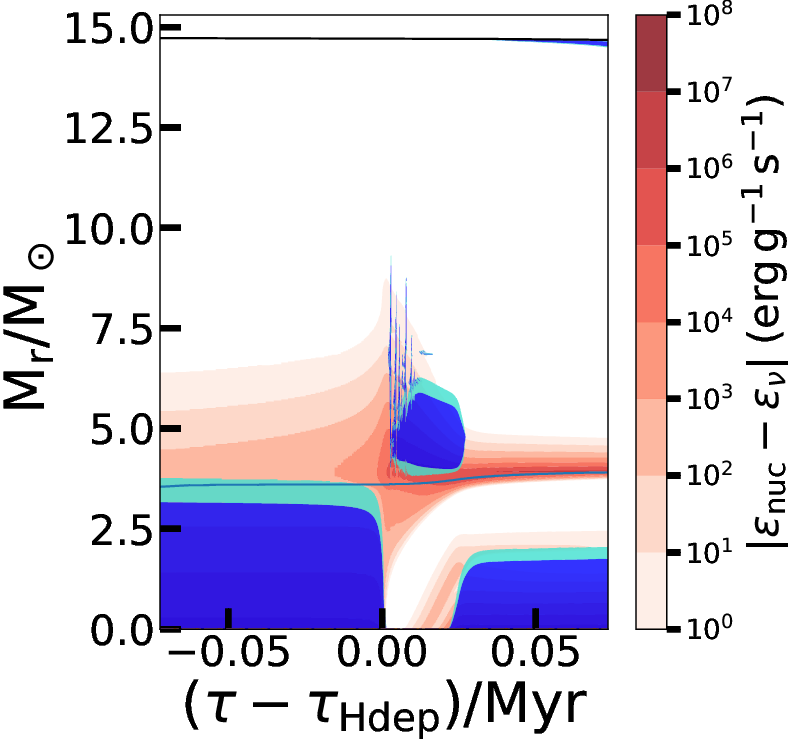}
	\hspace{0.2cm}
	\includegraphics[width=0.49\columnwidth]{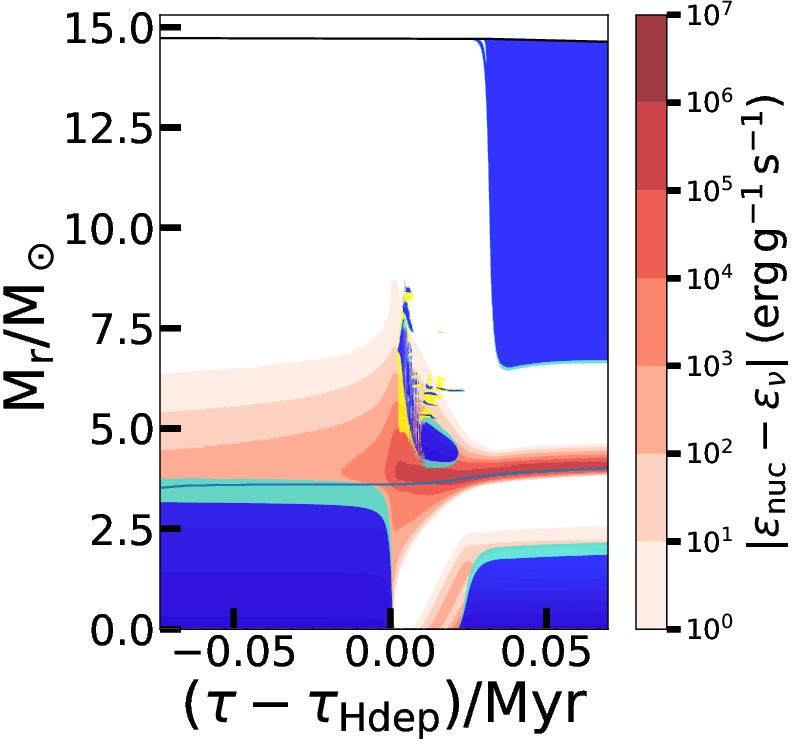}
	\caption{Structure evolution diagrams (Kippenhahn diagrams) of the $15\,\Msun$ models showing the ICZ. The left column presents the Schwarzschild model and the right column the Ledoux models with $\alphasc = 0.4$. The $\fcbm$ increases top to bottom with (0.0, 0.004, 0.01, 0.022). The blue region indicates convective regions, whereas the convective boundary region is shown in turquoise and semiconvection in the Ledoux models is shown in yellow. The red shading indicates the energy generation. The time on the x-axis is with respect to the time of core hydrogen depletion, $\tau_{\rm Hdep}$. The structure evolution diagrams are limited to the evolution between the locations where X$_c$($^1$H)$< 0.01$ and X$_c$($^4$He)$> 0.95$.}
	\label{kippenhahn_15M}
\end{figure}

\begin{figure}
	\includegraphics[width=0.49\columnwidth]{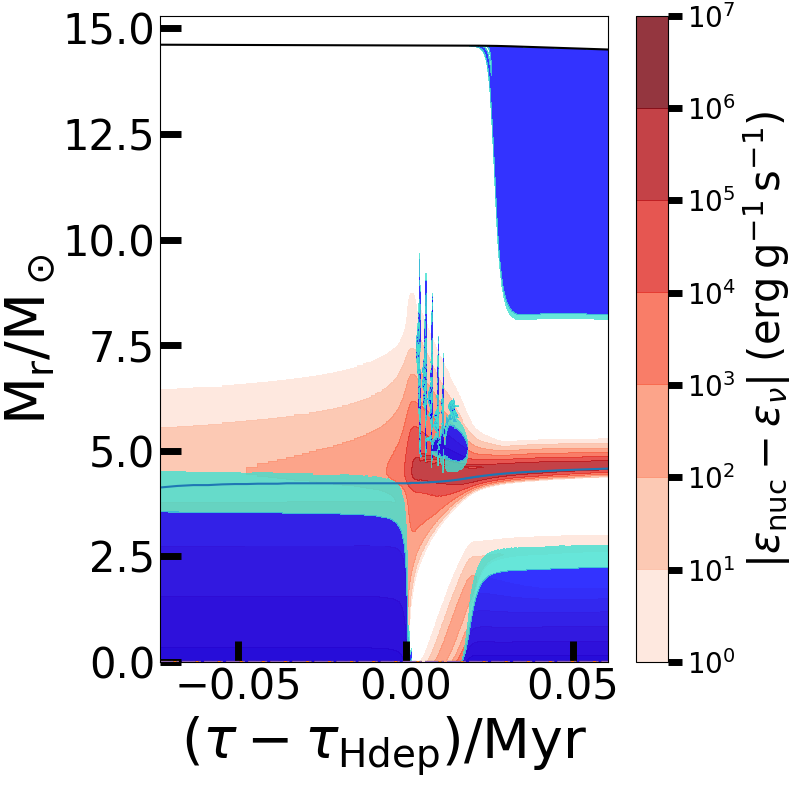}
	\hspace{0.2cm}
	\includegraphics[width=0.49\columnwidth]{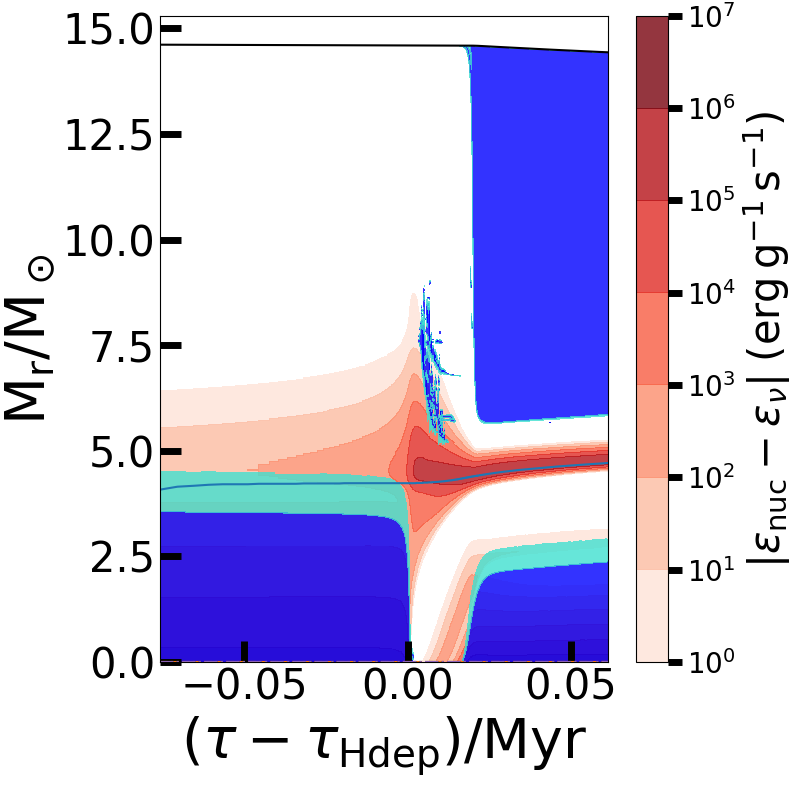}
	\includegraphics[width=0.49\columnwidth]{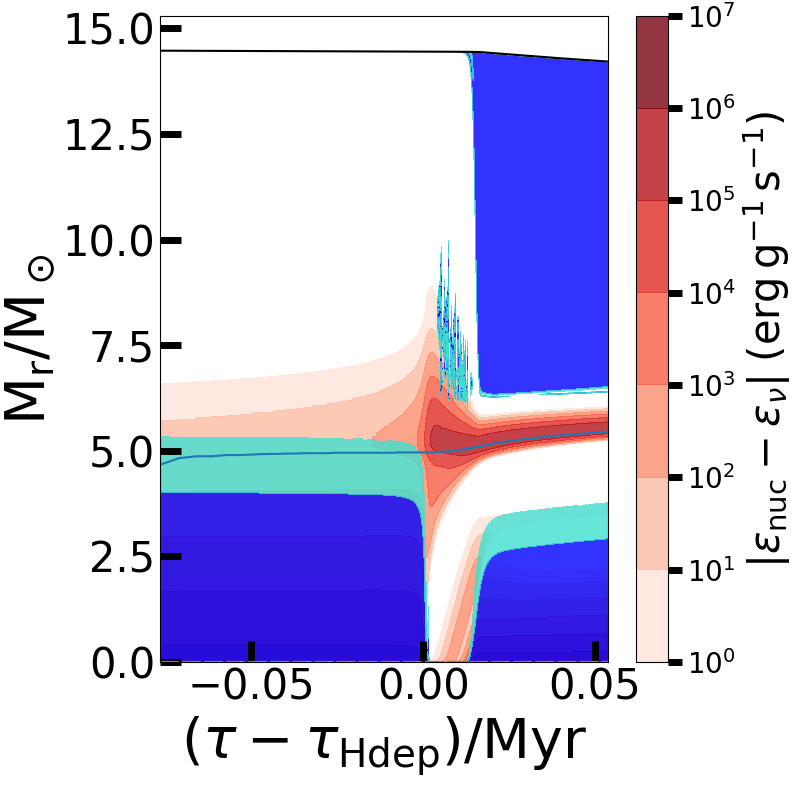}
	\hspace{0.2cm}
	\includegraphics[width=0.49\columnwidth]{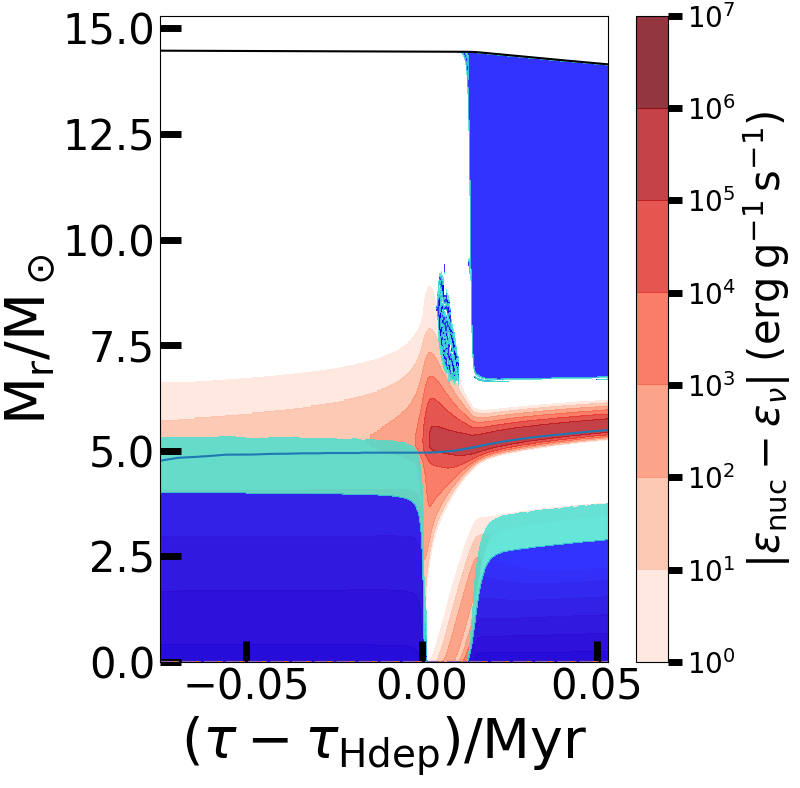}
	\caption{Same as Fig.~(\ref{kippenhahn_15M}) but for $\fcbm$ equal to 0.035 (top) and 0.05 (bottom).}
	\label{kippenhahn2_15M}
\end{figure}

After hydrogen is depleted in the core of the star, the convective core completely recedes and the star enters a short but crucial phase, which influences its fate. Since there is no nuclear energy generation left in the core, the star contracts, releasing gravitational energy. As a consequence of the virial theorem, energy conservation and a contraction on a short timescale the outer layer expands and cools down \citep[mirror principle - e.g.][]{Kippenhahn1994}. The layers above the previous hydrogen core, where there is still hydrogen left, heat up due to the contraction and set the condition for hydrogen burning. This hydrogen burning shell is accompanied by a convective layer, the intermediate convective zone (ICZ; Figs.~(\ref{kippenhahn_15M}), (\ref{kippenhahn2_15M}) and (\ref{kippenhahn_20M})).\\
It is during this phase that the star leaves the MS and, in the mass range studied here, crosses the HRD to the red super-giant (RSG) branch (Fig.~(\ref{HRD})). The details of this phase depend strongly on the duration, location and size of the ICZ with respect to the hydrogen burning shell. The properties of the ICZ, in turn, depend strongly on the choices of the convective boundary criterion and the amount of extra mixing at the boundary. If the ICZ only exists above the hydrogen burning shell, the latter can only consume the hydrogen at its location via nuclear burning and is consequently relatively weak. However, an overlap of the two creates a situation where the convective zone ingests fuel into the burning shell. This results in a much stronger burning shell which provides more support to the core against the gravitational pressure from the outer layers.\\
\begin{figure}
	\includegraphics[width=0.99\columnwidth]{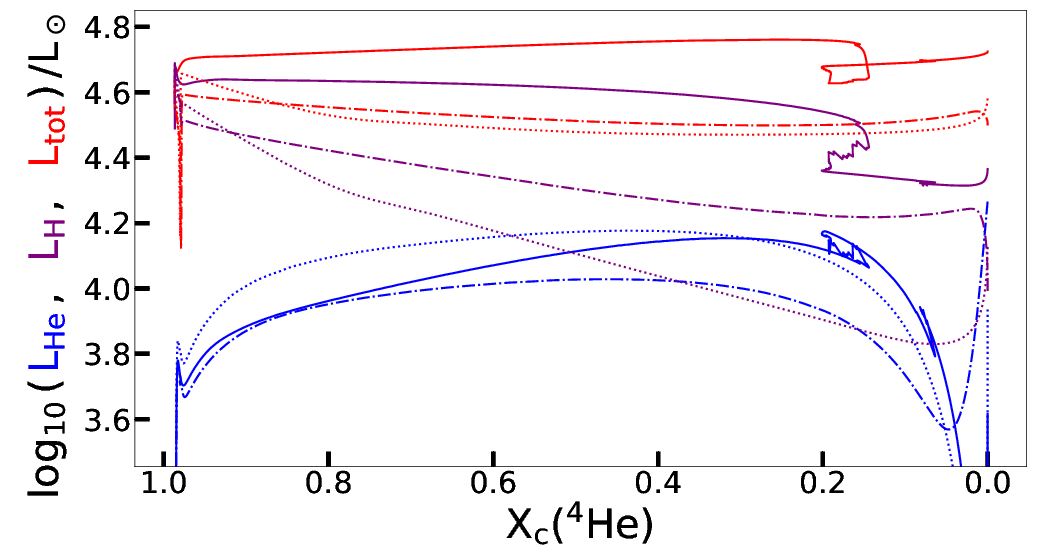}\\
	\includegraphics[width=0.99\columnwidth]{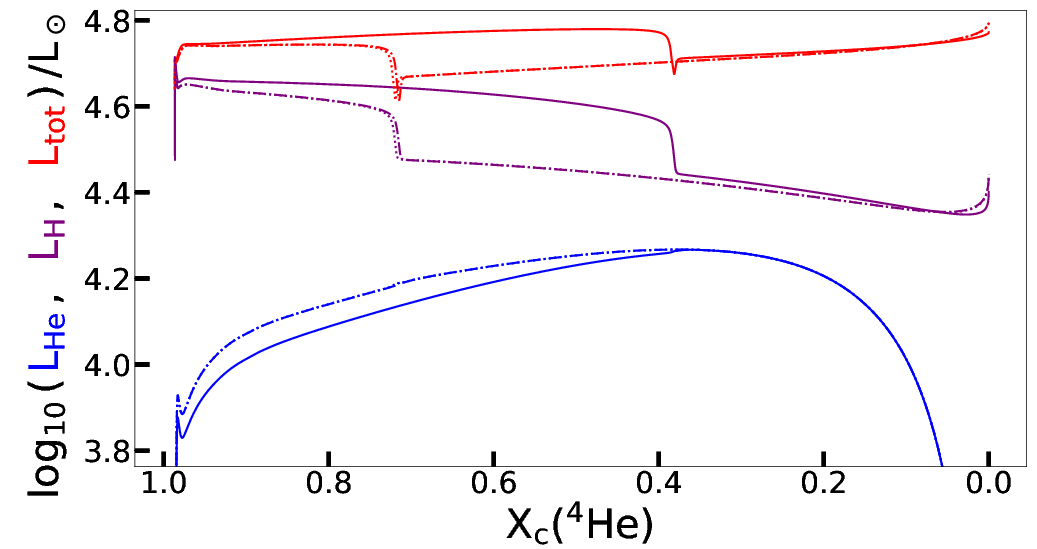}\\
	\includegraphics[width=0.99\columnwidth]{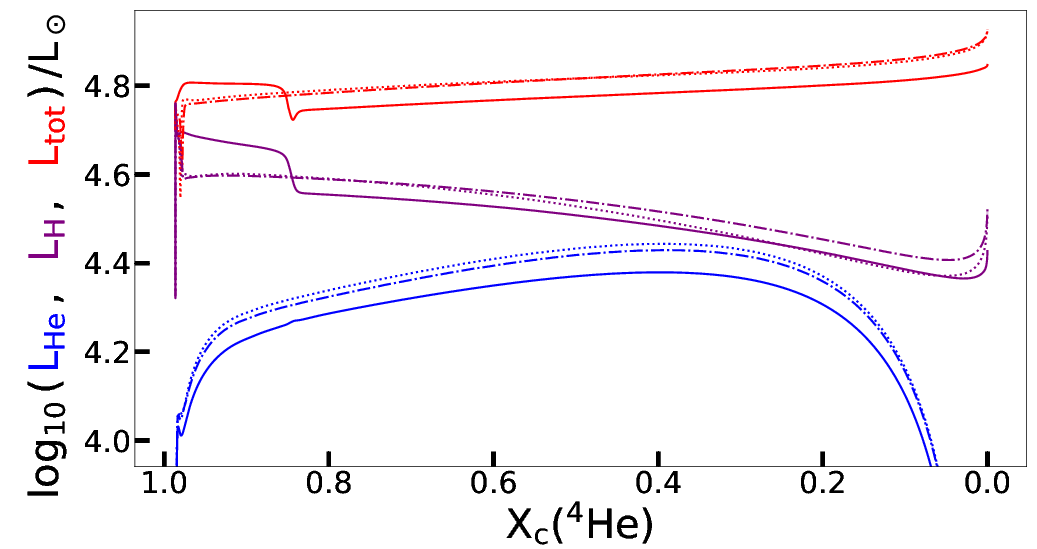}\\
	\includegraphics[width=0.99\columnwidth]{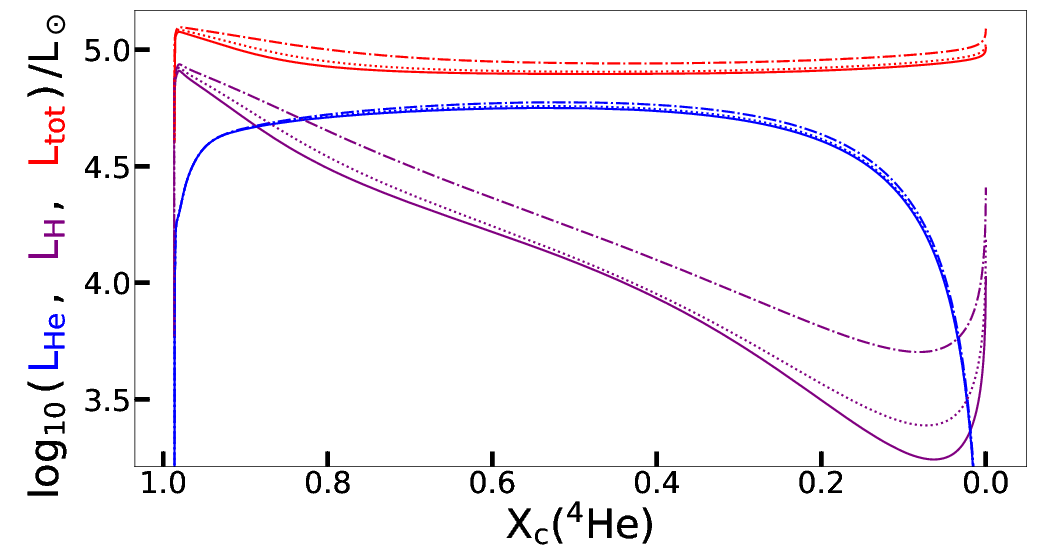}
	\caption{The total luminosity, $\rm{L}_{\rm tot}$ (red), and the luminosity generated by hydrogen and helium burning, $\rm{L}_{\rm H}$ (purple) and $\rm{L}_{\rm He}$ (blue) respectively, as a function of the central helium mass fraction. All the figures show $15\,\Msun$ models with varying $\fcbm = 0.0$, $0.01$, $0.022$ and $0.05$ (from top to bottom). Within one panel, all boundary criteria are shown, the \textit{Schwarzschild} (solid ) and the Ledoux one, the latter with $\alphasc = 0.004$ (dashdotted) and $\alphasc =0.4$ (dotted).}
		\label{luminosities_15M}
\end{figure}
Figs.~(\ref{kippenhahn_15M}) and (\ref{kippenhahn2_15M}) present structure evolution diagrams focussed on the ICZ. They show the amount of overlap between the ICZ and the hydrogen burning shell. Furthermore they visualise the size and give a hint of the duration of the ICZ. Fig.~(\ref{luminosities_15M}) presents the different post-MS luminosities of the simulations. Shown are the total luminosity and the luminosities generated by hydrogen and helium burning. The difference between the luminosities from the two burning types and the total luminosity is due to changes of the gravitational potential. The sudden drop in luminosity powered by hydrogen burning, if any, indicates the end of the boost of the ICZ, thus, its duration.\\
In the $15\,\Msun$ models with no CBM there is a clear difference between the Schwarzschild and the Ledoux models. The ICZ in the Schwarzschild model has an overlap with the hydrogen burning shell, whereas the Ledoux models develop an ICZ outside of the hydrogen burning shell. This difference arises because of the chemical composition gradient which prevents convection in the Ledoux models. These findings are similar to \citet{Langer1985}, \citet{Georgy2014} and \citet{Davies2018} who found that the depth at which the ICZ forms is sensitive to the stability criterion used.\\
The comparison between the Ledoux model with $\alphasc = 0.4$ and $\alphasc = 0.004$, both with no CBM, reveals that the ICZ appears at the same location. The small difference between the two is introduced by the mixing above the hydrogen core. Slow semiconvection is not able to remove the chemical composition gradient. Therefore, the intermediate convective region consists mainly a semiconvective region in the Ledoux model with slow semiconvection. In the case with efficient semiconvection, the \textit{convective fingers} partly removed the chemical composition gradient. Hence, the ICZ is mostly convective. This affects the time when the surface is enriched with hydrogen burning products, since the large surface convective zone penetrates into these layers shortly after the disappearance of the ICZ (Fig.~(\ref{kippenhahn_15M}), right upper corner). Furthermore, the energy transport in this region is more efficient when semiconvective layers, which only mix the chemical composition, are turned into convective layers. This slightly increases the luminosity as can be seen in Fig.~(\ref{luminosities_15M}), which in turn influences the mass loss rates. However, the impact is relatively small.\\
CBM changes this picture. The extra mixing at the boundary (i) removes possible chemical composition gradients at the boundary. Furthermore it increases the region with efficient mixing, hence, (ii) the energy excess is regulated faster and (iii) more fuel is provided for the burning shell. The latter simply increases the amount of boosting of the hydrogen shell. This is indicated by the hydrogen burning luminosity in Fig.~(\ref{luminosities_15M}), where the simulations with an overlap between the ICZ and the hydrogen shell have a higher L$_{\rm H}$ for the duration of the ICZ before the hydrogen powered luminosity drops. The second point decreases the lifetime of the intermediate convective core by increasing the region with efficient energy transport, thus, $\gradr$ drops faster. This is illustrated in Figs.~(\ref{kippenhahn_15M}) and (\ref{kippenhahn2_15M}), where the models with larger values of $\fcbm$ have a shorter duration of the ICZ. Furthermore, in Fig.~(\ref{luminosities_15M}) the luminosity powered by the hydrogen burning shell experiences the drop earlier with higher $\fcbm$. In the most extreme cases with $\fcbm = 0.05$ and the Ledoux models with $\fcbm = 0.035$ the envelope to core ratio is too small to produce a proper ICZ that never overlaps with the hydrogen shell (Fig.~\ref{kippenhahn2_15M}). In these models, L$_{\rm H}$ in Fig.~(\ref{luminosities_15M}) constantly drops, very similar to the Ledoux models with no CBM. The first point crucially impacts the Ledoux models, because it efficiently removes the $\mu$-gradient at the convective boundary, which prevents the ICZ from moving inward. Consequently, the ICZ moves downwards in mass coordinates and eventually\footnote{The downward movement is not instantaneous because only the $\mu$-gradient in the convective boundary layer is erased. Hence, the overlap of the ICZ and the hydrogen shell in the Ledoux models, if any, always is delayed compared to the Schwarzschild models (compare left and right column in Fig.~(\ref{kippenhahn_15M})).} overlaps with the hydrogen burning shell (Fig.~(\ref{kippenhahn_15M}), right column). In the Ledoux models there always is a short semiconvective region before the ICZ penetrates downward. However, semiconvection is not efficient enough for the $\alphasc$ values tested in this work to erase the chemical composition gradient by themselves because of the short timescale of this evolutionary phase. Moreover, when moving downwards the ICZ leaves behind a chemical composition gradient. Therefore, the ICZ has, when it starts boosting the burning shell, a semiconvective zone at its upper boundary (Fig.~(\ref{kippenhahn_15M})). These semiconvective regions, however, become smaller as $\fcbm$ is increased and disappear for the two largest values used. Contrary, the ICZ in the Schwarzschild models include this region, hence, they span a wider region and are able to boost the hydrogen shell for a longer time. This creates the difference in the luminosity powered by hydrogen burning between the \textit{Ledoux} and \textit{Schwarzschild} criterion in Fig.~(\ref{luminosities_15M}).\\
It should be noted, that the impact of the above mentioned points (ii) and (iii) affect the ICZ differently; (ii) reduces the duration of the convective shell, whereas (iii) boosts the hydrogen burning region more, which in turn leads to a longer duration of the ICZ. In the Schwarzschild model (ii) leads to a decrease of the duration of the ICZ (Figs.~(\ref{kippenhahn_15M}), (\ref{kippenhahn2_15M}) and (\ref{luminosities_15M})). In the Ledoux models, on the other hand, at low $\fcbm$ (i) dominates. This leads to a boost of the ICZ due to the ingestion of fuel into the burning shell. When increasing the amount of CBM, the point (ii) starts to reduce the duration of the ICZ, similar to the Schwarzschild models.\\
CBM does not change the initial location of the ICZ. The \texttt{Ledoux} criterion always predicts the initial location above the hydrogen shell, whereas the \texttt{Schwarzschild} criterion always predicts an overlap (Fig.~(\ref{kippenhahn_15M}), except when the envelope to core ratio is too small to produce a ICZ as in the models with $\fcbm = 0.05$). This difference, and point (ii) above, also lead to the two behaviours, overlap and no overlap, in the $15\,\Msun$ models with $\fcbm = 0.035$ in Fig.~(\ref{kippenhahn2_15M}), where the ICZ in the Ledoux model does not overlap but in the Schwarzschild model it does.\\
\citet{Davies2018} predict the first overlap of the ICZ and the hydrogen burning shell in their Ledoux models around $16 M_\odot$. We show here that the lowest initial mass that shows an overlap is dependent on the amount of mixing at the convective boundary. Furthermore, an overall result is that the differences of the ICZ in the $15\,\Msun$ models due to the choice of the stability criterion decrease with increasing amount of CBM, and for $\fcbm = 0.05$ Figs.~(\ref{kippenhahn2_15M}) and (\ref{luminosities_15M}) show very similar results.\\
\begin{figure}
	\includegraphics[width=0.49\columnwidth]{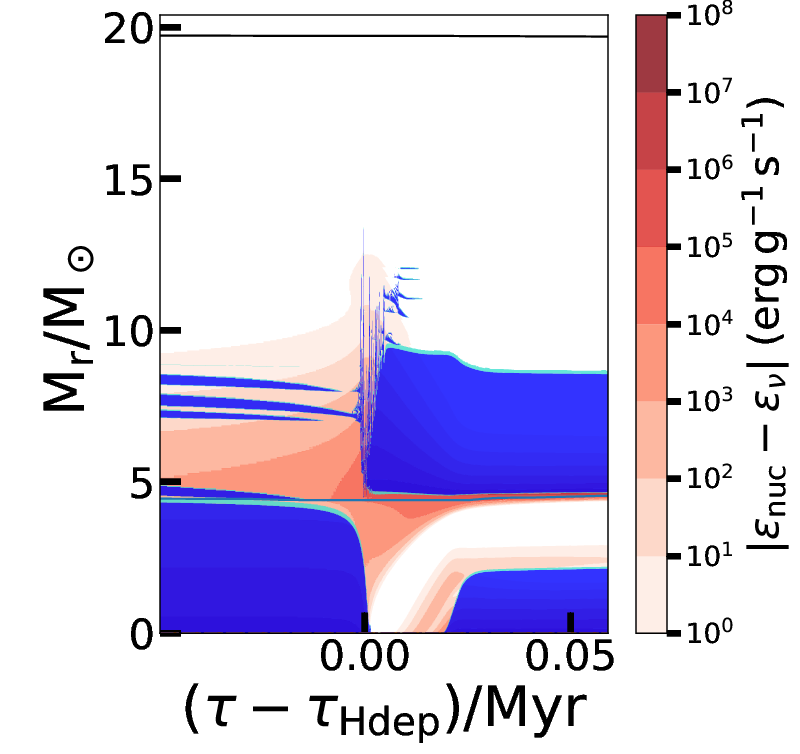}
	\hspace{0.2cm}
	\includegraphics[width=0.49\columnwidth]{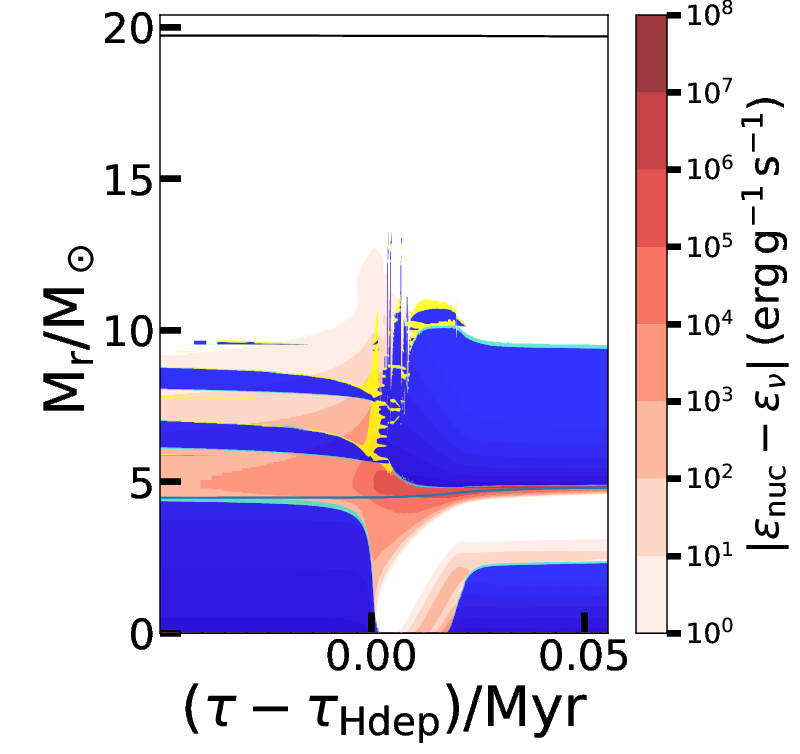}\\
	\includegraphics[width=0.49\columnwidth]{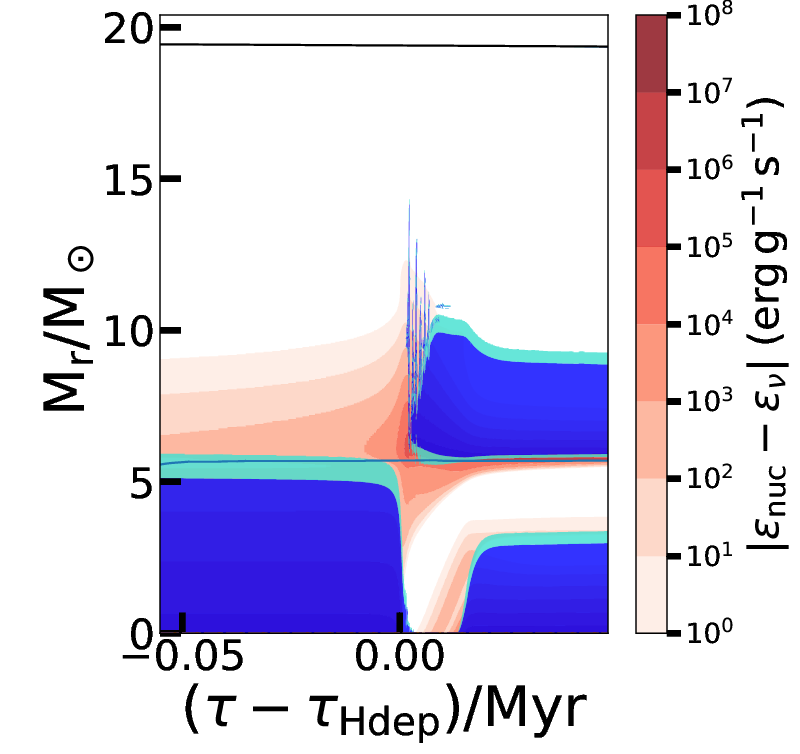}
	\hspace{0.2cm}
	\includegraphics[width=0.49\columnwidth]{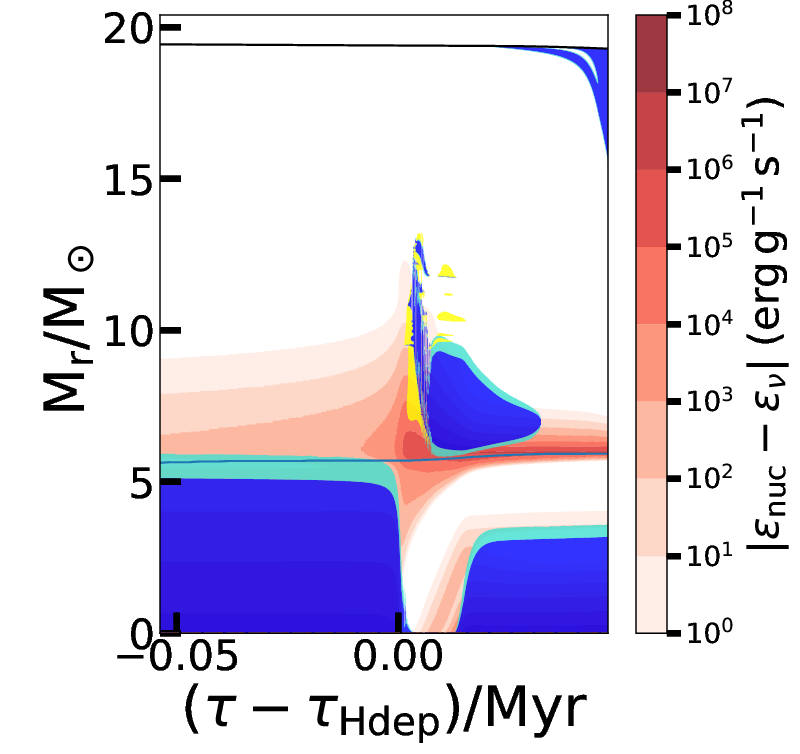}\\
	\includegraphics[width=0.49\columnwidth]{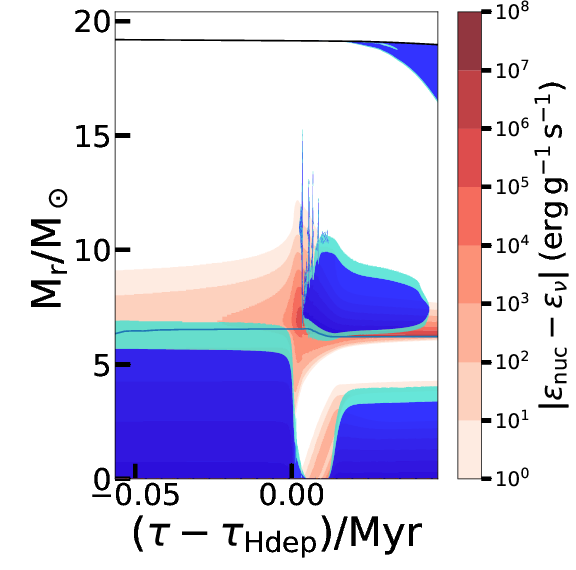}
	\hspace{0.2cm}
	\includegraphics[width=0.49\columnwidth]{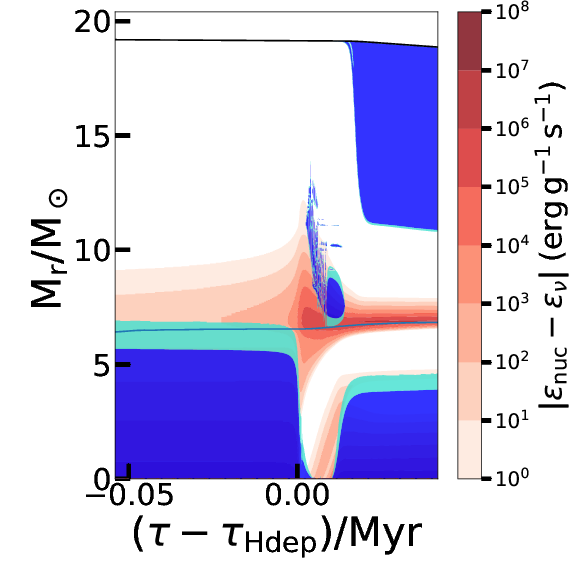}
	\caption{Structure evolution diagram of the $20\,\Msun$ simulations showing the ICZ as in Fig.~(\ref{kippenhahn_15M}). The left column presents the Schwarzschild model and the right column the Ledoux models with $\alphasc = 0.4$. The top row uses $\fcbm = 0.004$, the middle $\fcbm = 0.022$ and the bottom $\fcbm = 0.035$.}
	\label{kippenhahn_20M}
\end{figure}
The ICZ exhibits the same dependence on CBM in the $20$ and $25\,\Msun$ models  as in the $15\,\Msun$ models. There are, however, some important differences. As discussed in Section \ref{HydrogenBurning}, in stars with higher initial masses (a) the luminosity is higher and (b) the \textit{convective fingers} above the convective core are more present. (a) leads to an increased radiative temperature gradient in the region above the hydrogen burning shell. Consequently, the ICZ spans a larger radial distance in the models with higher initial masses (e.g. Fig.~(\ref{kippenhahn_20M})). Therefore, more fuel is provided for the hydrogen burning shell and it is boosted longer. This in turn prolongs the lifetime of the ICZ, leading to an ICZ that can be present during nearly all of the core helium burning lifetime (see Table \ref{properties_hedep}). In general, the relative duration of the ICZ with respect to the core helium burning duration increases with initial mass and, in accordance to the previous discussion, decreases with $\fcbm$. (b) may lead to \textit{convective fingers} that exist until the appearance of the ICZ (e.g. Fig.~(\ref{kippenhahn_20M}), top row). These convective layers partly remove the chemical composition profile left behind by the receding convective hydrogen core. This mainly influences the Ledoux models, where the ICZ overlaps much faster (nearly at the same time as in the Schwarzschild model, Fig.~(\ref{kippenhahn_20M})). Moreover, the ICZ in the Ledoux model is slightly bigger compared to the Schwarzschild model because of the slightly higher temperature at the location of the hydrogen shell. Therefore it can replenish the hydrogen shell with fuel for longer and is active for longer compared to the Schwarzschild models of the same initial mass. Therefore the ICZ last longer in the Ledoux models than in the Schwarzschild models with higher initial masses.\\
In the $20\,\Msun$ with $\fcbm = 0.022$ there are no \textit{convective fingers} at the same mass coordination where the ICZ eventually appears. Therefore, the Ledoux model behaves very similar to the $15\,\Msun$ model. As a result, the drop in luminosities of the Ledoux model are much earlier compared to the models with less CBM.\\
The $25 M_\odot$ Ledoux model with $f = 0.01$ breaks out of the general trend by creating an ICZ which lasts longer than the core helium burning (similar to \citet{Ritter2018}, their Fig.(11)).\\
The different behaviour in depth and duration of the ICZ has an important impact on the further evolution of the star. In summary, the ICZ influences the strength of the hydrogen burning shell. This shell is crucial in determining the further evolution since it supports the contracting core underneath against the gravitational pressure from the outer layers, which affect the way the star evolves through this short phase and hence sets the structure for its further evolution, e.g. the convective helium core (Section \ref{heliumBurning}) or the surface evolution (Section \ref{BSG_vs_RSG}).

\section{Core Helium Burning} \label{heliumBurning}

\subsection{Convective Helium Core} \label{ConvectiveHeliumCore}

During the helium burning stage, the convective helium core constantly grows in mass. This is because (i) the increase of the core luminosity due to the active hydrogen burning shell which continuously synthesises hydrogen into helium, thus, increasing the helium core mass, (ii) the increase of opacity and mean molecular weight due to the conversion of helium into carbon and oxygen and (iii) the density dependence of the $3\alpha$ (second order) and $^{12}$C($\alpha,\gamma$)$^{16}$O (first order) reaction rate.\\
\begin{figure}
	\includegraphics[width=\columnwidth]{figures/legend_convHydrogenCore_15M_f00p002_colorblind_revision.png}
	\includegraphics[width=\columnwidth]{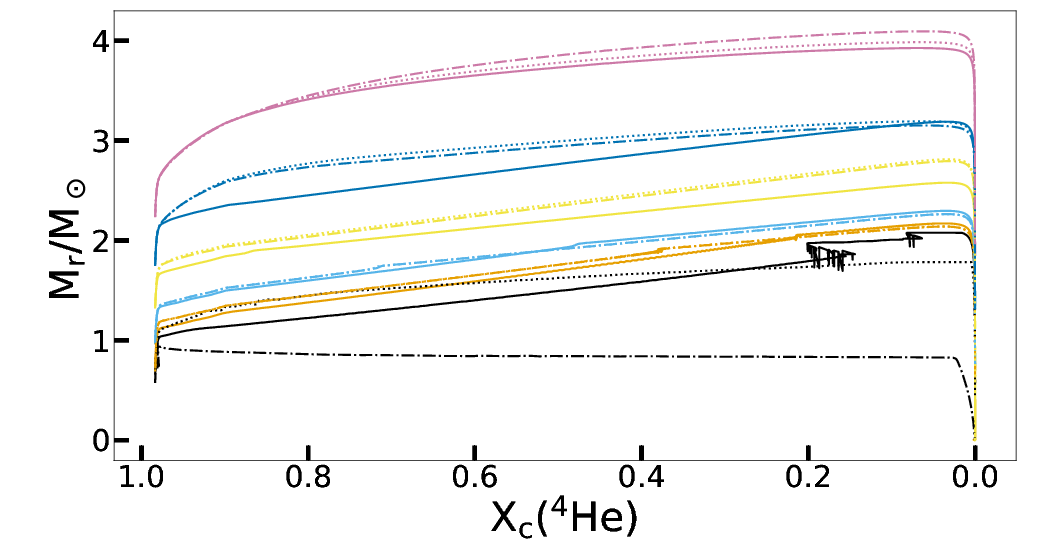}\\
	\includegraphics[width=\columnwidth]{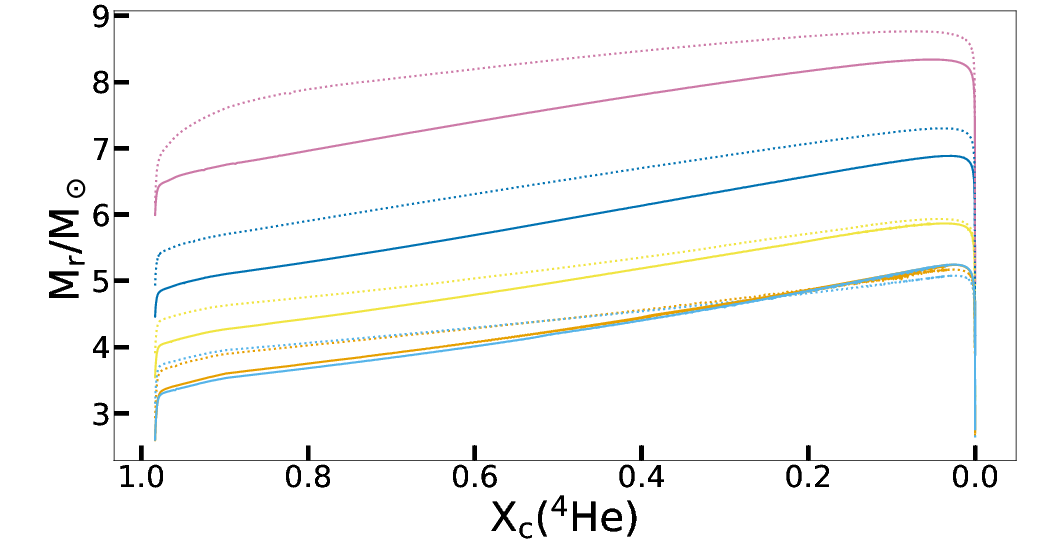}
	\caption{The location of the convective helium core boundary, determined either by the \textit{Ledoux} or the \textit{Schwarzschild} criterion, as a function of the central helium mass fraction. Shown are the $15$ (top) and $25\,\Msun$ (bottom) models. The boundary criterion is shown by the linestyle, where a solid line indicates the Schwarzschild criterion and the Ledoux criterion shown with a dashdotted ($\alpha_{\rm{sc}} = 0.004$) or a dotted line ($\alpha_{\rm{sc}} = 0.4$). The color scheme is the same as in Fig.~(\ref{boundaryLocation_hydrogenCore})}
	\label{convHeliumCoreMass}
\end{figure}
Fig.~(\ref{convHeliumCoreMass}) presents the location of the convective helium core boundary as a function of the central helium mass fraction. The boundary shown is the convective core determined by the stability criterion without the boundary mixing region. The size and growth of the convective helium core depends on (i) the amount of mixing at the convective boundary, (ii) the strength and location of the hydrogen shell (Section \ref{ICZ}) and (iii) on the choice of the stability criterion. Fig.~(\ref{convHeliumCoreMass}) clearly illustrates that the convective core is larger in the models with more CBM for a given convective boundary criterion. It furthermore shows that the different sizes of the convective cores arise mainly during their initial growth. During the rest of the core helium burning phase the cores grow at a similar rate. Interestingly, the models applying the \textit{Ledoux} criterion predict a faster initial growth of the convective core than the corresponding Schwarzschild models, with the exception of the models with no CBM and the models with $\fcbm = 0.05$. In the latter the Schwarzschild model initially predicts a convective helium core which is very slightly larger before the helium core in the Ledoux models overtake it (Fig.~\ref{convHeliumCoreMass}). The initial size of the convective core depends strongly on the activity of the hydrogen shell, which itself is strongly affected by the ICZ (see Section \ref{ICZ}). A stronger hydrogen shell supports the core against the gravitational pull from the outer layers. Consequently, the helium core contracts less and the burning is slightly less energetic, hence, because of $\nabla_{\rm{rad}} \propto \ell$, the convective helium core is smaller.\\
A larger amount of CBM increases the mixed zone above the convective core and smooths the chemical composition gradient at the boundary. The first point provides the central burning region with fuel, increasing the energy generation. This results in a higher luminosity (L$_{\rm He}$ in Fig.~(\ref{luminosities_15M})), thus a larger convective core. The second point removes the limiting chemical composition gradient in the Ledoux models. Consequently, the growth of the convective core in these models is less limited than in the models with no CBM. Additionally, the semiconvective regions above the convective core disappear in the models with CBM because $\mu$-gradient is only non-zero in the radiative layers. Therefore, the relative importance of semiconvection on the evolution of the convective helium core in the simulation with CBM is reduced. The small differences between the Ledoux model with CBM and different semiconvective efficiency arise from the different strength of the hydrogen burning shell (Section \ref{ICZ}).\\
After its initial growth the convective core mass continues to increase because of the growth of the helium core mass due to the active hydrogen shell and the increase in opacity. This nearly constant increase is occasionally disrupted by kinks (e.g. in the Schwarzschild model with $\fcbm = 0.022$ at $\rm{X}_c(^4\rm{He}) \sim 0.87$ in Fig.~(\ref{convHeliumCoreMass})), which are a consequence of the convective core regulating  itself to the changes of the energy generation in the hydrogen shell or the opacity in the core.\\
The kinks around  X$_c$($^4$He) $\sim 0.7$ and $0.5$ for the models with $f = 0.01$, around  X$_c$($^4$He) $\sim 0.35$ for the models with $f = 0.004$ occur due to vibrational up-down movements of the convective boundary. This ingests a higher amount of fuel into the convective zone and finally increases the zone as a consequence of the higher energy generation. We omit the discussion here, whether these are \textit{core breathing pulses} \citep{Castellani1985} or numerical artefacts \citep[e.g.][]{Constantino2016, Farmer2016} and call these events \textit{core breathing pulses} out of convenience. The presence of the \textit{core breathing pulses} is discussed in Section \ref{discussion}. However, we want to outline that the intensity of these \textit{core breathing pulses} decreases, or they even vanish, with increasing amount of CBM, e.g. $\fcbm \geqslant 0.022$ for the $15\,\Msun$ models. The messy behaviour of the $15\,\Msun$ Schwarzschild model with no CBM is due to a fast, nearly step-like, increase of the convective core.\\
The differences in the convective core size of the $15\,\Msun$ models with the same amount of CBM but different boundary criteria, which are more dominant for $\fcbm \gtrsim 0.022$, arise because of the different (i) amounts of energy generation and (ii) radial location of the hydrogen burning shell. (i) supports the core more or less against the gravitational pressure of the outer layers, where a higher energy output by the burning shell leads to a smaller convective core. (ii), on the other hand, changes the helium core mass. If the burning shell is further out or moves outwards faster (due to a smaller amount of fuel available), the core mass is bigger, hence a higher helium burning luminosity (Fig.~\ref{luminosities_15M}) and a larger convective core. This dependency is apparent when comparing Figs.~(\ref{kippenhahn_15M}), (\ref{kippenhahn2_15M}), (\ref{luminosities_15M}) and (\ref{convHeliumCoreMass}). The behaviour is not linear because the interaction between the ICZ and the hydrogen burning region is not linear (see Section \ref{ICZ}). Therefore, contrary to the trend of finding convergence between the two convective stability criteria with an increasing amount of CBM during core hydrogen burning, the different sizes of the convective helium core between the Schwarzschild and the Ledoux models varies more for $\fcbm \gtrsim 0.022$ (apart from the case with no CBM).\\
These uncertainties of the convective helium core affects the helium and carbon-oxygen core masses, see also Table \ref{properties_hedep}. This will influence the further central evolution and affect the pre-supernovae structure, which depends on these core masses.\\
In the Ledoux models with no CBM the convective core grows to a size of about $1\,M_\odot$ (Fig.~(\ref{convHeliumCoreMass}), black dashed and dashdotted lines) before a chemical composition gradient is built up above the convective core. The radiative temperature gradient continues to increase and a semiconvective region develops above the core. The semiconvection in this model is not efficient enough to fully remove the restricting $\mu$-gradient built up by the convective core and the convective helium core stops growing for the rest of this burning phase. Above this core, however, several sandwiched layers of semiconvection and convection occur, which increase in number with time, because the semiconvective process slowly erases the $\mu$-gradient\footnote{When the $\mu$-gradient decreases the term $(\nabla_{\rm L} - \nabla)^{-1}$ in Eq.~(\ref{semiconvectionFromulae}) increases, which enhances the semiconvective mixing in this layer. The result is a very spiky chemical composition profile.}. In the model with fast semiconvection, the chemical composition gradient is steadily removed by a thin semiconvective layer just above the convective core. This, however, leads to a wiggly convective core boundary but the core can initially grow very similarly to the convective core in the Schwarzschild model. The Schwarzschild model with no CBM ignores the $\mu$-gradient and initially grows similarly to the other Schwarzschild models with CBM. At around X$_c$($^4$He)$\approx 0.9$ the convective core growth plateaus before it continues to grow further at X$_c$($^4$He)$\approx 0.8$. This is a result of the hydrogen shell, which is boosted there in the Schwarzschild model as a result of the interaction with the ICZ. Consequently, the Ledoux model with fast semiconvection predicts a bigger convective helium core at the beginning of core helium burning. As the evolution proceeds, however, a chemical composition gradient builds up above the core, which becomes too strong for semiconvection to erase. This reduces the increase of the convective core. The convective core in the Schwarzschild model on the other hand grows further, predicting a overall larger convective helium core than in the Ledoux models (Table \ref{properties_hedep}).\\
The chaotic behaviour of the core boundary around X$_c$($^4$He)$\approx 0.1$ in the Schwarzschild model with no CBM (solid red line in Fig.~(\ref{convHeliumCoreMass})) is due to a convective pillar that rises on top of the convective core, much stronger than the \textit{core breathing pulses} previously mentioned. We tested this behaviour against an increased resolution but the feature remained.\\ 
The lower panel in Fig.~(\ref{convHeliumCoreMass}) shows the convective helium core boundary of the various $25\,\Msun$ models. The convective helium core grows with time as in the $15\,\Msun$ models but there are some important differences, which are more prominent in the models with larger $\fcbm$. These differences, apart from the generally larger convective helium core with increasing initial mass, are due to the different behaviour of the ICZ.\\
In the $25\,M_\odot$ case, the initial growth of the convective helium core is larger in all Ledoux models than in the Schwarzschild models with the same amount of CBM. This is because the hydrogen shell is less active in the latter and slightly closer to the convective core. This is a consequence of the different evolution of the ICZ similar to Fig.~(\ref{kippenhahn_20M}), middle and bottom row. Therefore, the Ledoux models predict a larger convective helium core than the Schwarzschild models with the same amount of convective boundary mixing. The gap is larger for higher $\fcbm$ values. Interestingly, the behaviour of the ICZ in the $25\,M_\odot$ models leads to convective helium cores more similar in the Schwarzschild with $f = 0.004$ and $0.01$ than with their Ledoux counterparts (and vice-versa for the Ledoux models, Fig.~(\ref{convHeliumCoreMass} and Table \ref{properties_hedep})). The convective core in the Schwarzschild models experience a faster growth starting around X$_c$($^4$He) $\sim 0.5$. There the hydrogen shell narrows and with it the ICZ. Consequently the core generates more energy and the convective region grows. This finally leads to similar core masses at the end of core helium burning for all the $25\,\Msun$ models with $f = 0.004$, $0.01$ and $0.022$.\\
The behaviour of the convective helium core in the $20\,\Msun$ models is a mixture of the behaviours of the two other initial masses. The simulations with $f = 0.004$ behave similarly to the $25\,M_\odot$ case with the exception of the stronger \textit{core breathing pulse}. The calculations with $f = 0.022$ are similar to the $15\,M_\odot$ models with the exception that the hydrogen shell are at about the same location in the Ledoux and Schwarzschild models. Therefore the convective core size grows at a similar rate, apart from an increase around X$_c$($^4$He)$\sim 0.6$, which is due to the disappearance of the ICZ (this is no \textit{core breathing pulse}). The $20\,\Msun$ models with the two largest values of $\fcbm$ show a different ICZ and hence different convective helium cores depending on the boundary criterion, similar to the $25\,\Msun$ models.\\
The convective core in the $20\,\Msun$ models with $f = 0.01$ are the exception from the above discussion. There the Schwarzschild model predict a larger convective helium core than the Ledoux models. This is because of the larger ICZ in the Ledoux models compared to the Schwarzschild models. Therefore, the Schwarzschild model provides more energy from central helium burning compared to the Ledoux model, hence, the relatively smaller convective core in the latter.
\begin{table*}
	\centering
	\begin{threeparttable}
	\caption{Properties of the stellar models at core helium depletion.}
	\begin{tabular}{llcccccccc}\\
\toprule
		model  & $\fcbm$ \& $\alphasc$ & M$_{\rm{final}}$ & $\rm{M}_\alpha$\tnote{a} & $\rm{M}_\alpha$ & M$_{\rm{CO}}$ & $\tau_{\rm{H}}$ & $\tau_{\rm He}$ & $\tau_{\rm BSG}/\tau_{\rm He}$\tnote{b} & log$_{10}\,\rm{T}_{\rm eff, min}^{\rm MS}$\tnote{c} \\
		 &   & (M$_\odot$) & (M$_\odot$) & (M$_\odot$) & (M$_\odot$) & (Myrs) & (Myrs) &  & (K) \\
\midrule
		$15 \rm{M}_\odot$, & $\fcbm = 0.0$, & 14.35 & 2.55 & 4.07 & 2.08 & 11.08 & 1.47 & 0.78 & 4.40 \\
		Schwarzschild, & $\fcbm = 0.004$ & 13.83 & 2.73 & 4.08 & 2.14 & 11.40 & 1.48 & 0.54 & 4.39 \\
		$f_0 = 0.002$ & $\fcbm = 0.01$ & 13.43 & 3.03 & 4.28 & 2.34 & 11.93 & 1.34 & 0.36 & 4.38 \\
		 & $\fcbm = 0.022$ & 12.34 & 3.60 & 4.68 & 2.75 & 12.86 & 1.19 & 0.02 & 4.36 \\
		 & $\fcbm = 0.035$ & 11.23 & 4.23 & 5.39 & 3.45 & 13.74 & 1.03 & 0.02 & 4.32 \\
		 & $\fcbm = 0.05$ & 11.13 & 4.95 & 6.22 & 4.33 & 14.60 & 0.84 & 0.01 & 4.28 \\
\hline
		$15 \rm{M}_\odot$, & $\fcbm = 0.004$ & 14.28 & 2.65 & 4.05 & 2.10 & 11.28 & 1.52 & 0.77 & 4.40 \\
		Schwarzschild, & $\fcbm = 0.01$ & 13.61 & 2.96 & 4.24 & 2.29 & 11.83 & 1.39 & 0.46 & 4.38 \\
		$f_0 = 0.02$  & $\fcbm = 0.022$ & 12.47 & 3.59 & 4.65 & 2.72 & 12.81 & 1.20 & 0.02 & 4.36 \\
\hline
		$15 \rm{M}_\odot$, & $\fcbm = 0.0$, $\alpha_{\rm{sc}} = 0.004$ & 14.30 & 2.21 & 3.51 & 0.56 & 10.47 & 0.93 & 0.02 & 4.41 \\
		Ledoux  & $\fcbm = 0.0$, $\alpha_{\rm{sc}} = 0.4$ & 14.09 & 2.54 & 3.73 & 2.13 & 11.07 & 1.26 & 0.01 & 4.40 \\
		$f_0 = 0.002$  & $\fcbm = 0.004$, $\alpha_{\rm{sc}} = 0.004$ & 13.85 & 2.72 & 4.07 & 2.13 & 11.40 & 1.44 & 0.49 & 4.39 \\
		  & $\fcbm = 0.004$, $\alpha_{\rm{sc}} = 0.4$ & 13.85 & 2.72 & 4.07 & 2.13 & 11.40 & 1.44 & 0.49 & 4.39 \\
		  & $\fcbm = 0.01$, $\alpha_{\rm{sc}} = 0.004$ & 13.18 & 3.02 & 4.24 & 2.31 & 11.93 & 1.34 & 0.15 & 4.38 \\
		  & $\fcbm = 0.01$, $\alpha_{\rm{sc}} = 0.4$ & 13.18 & 3.02 & 4.24 & 2.31 & 11.93 & 1.33 & 0.15 & 4.38 \\
		  & $\fcbm = 0.022$, $\alpha_{\rm{sc}} = 0.004$ & 11.88 & 3.60 & 4.92 & 2.94 & 12.86 & 1.16 & 0.01 & 4.36 \\
		  & $\fcbm = 0.022$, $\alpha_{\rm{sc}} = 0.4$ & 11.92 & 3.60 & 4.96 & 2.98 & 12.86 & 1.14 & 0.01 & 4.36 \\
		  & $\fcbm = 0.035$, $\alpha_{\rm{sc}} = 0.004$ & 12.46 & 4.23 & 5.30 & 3.40 & 13.74 & 0.95 & 0.02 & 4.32 \\
		  & $\fcbm = 0.035$, $\alpha_{\rm{sc}} = 0.4$ & 12.41 & 4.23 & 5.34 & 3.44 & 13.74 & 0.94 & 0.02 & 4.32 \\
		  & $\fcbm = 0.05$, $\alpha_{\rm{sc}} = 0.004$ & 10.39 & 4.96 & 6.41 & 4.50 & 14.60 & 0.84 & 0.01 & 4.28 \\
		  & $\fcbm = 0.05$, $\alpha_{\rm{sc}} = 0.4$ & 10.90 & 4.95 & 6.28 & 4.39 & 14.60 & 0.83 & 0.01 & 4.28 \\
\hline
		$15 \rm{M}_\odot$, & $\fcbm = 0.004$, $\alpha_{\rm{sc}} = 0.004$ & 13.63 & 2.64 & 4.06 & 2.14 & 11.25 & 1.27 & 0.01 & 4.40 \\
		Ledoux, & $\fcbm = 0.01$, $\alpha_{\rm{sc}} = 0.004$ & 13.40 & 2.96 & 4.20 & 2.27 & 11.82 & 1.35 & 0.27 & 4.38 \\
		$f_0 = 0.02$ & $\fcbm = 0.022$, $\alpha_{\rm{sc}} = 0.004$ & 11.90 & 3.59 & 4.91 & 2.94 & 12.82 & 1.16 & 0.01 & 4.36 \\
\hline
		$20 \rm{M}_\odot$, & $\fcbm = 0.004$ & 17.37 & 4.39 & 5.94 & 3.63 & 8.12 & 0.99 & 0.64 & 4.43 \\
		Schwarzschild, & $\fcbm = 0.01$ & 17.49 & 4.89 & 6.24 & 3.95 & 8.40 & 0.89 & 0.64 & 4.42 \\
		$f_0 = 0.002$ & $\fcbm = 0.022$ & 14.48 & 5.70 & 6.71 & 4.45 & 8.95 & 0.82 & 0.17 & 4.38 \\
		  & $\fcbm = 0.035$ & 11.93 & 6.54 & 7.35 & 5.13 & 9.46 & 0.78 & 0.02 & 4.34 \\
		  & $\fcbm = 0.05$ & 10.95 & 7.49 & 8.74 & 6.52 & 9.97 & 0.66 & 0.01 & 4.27 \\
\hline
		$20 \rm{M}_\odot$, & $\fcbm = 0.004$, $\alpha_{\rm{sc}} = 0.4$ & 18.90 & 4.49 & 5.89 & 3.58 & 8.09 & 0.96 & 0.92 & 4.43 \\
		Ledoux, & $\fcbm = 0.01$, $\alpha_{\rm{sc}} = 0.4$ & 18.66 & 4.90 & 6.13 & 3.86 & 8.40 & 0.92 & 0.87 & 4.42 \\
		$f_0 = 0.002$ & $\fcbm = 0.022$, $\alpha_{\rm{sc}} = 0.4$ & 13.02 & 5.70 & 6.90 & 4.61 & 8.95 & 0.79 & 0.02 & 4.38 \\
		  & $\fcbm = 0.035$, $\alpha_{\rm{sc}} = 0.4$ & 10.84 & 6.54 & 8.01 & 5.71 & 9.46 & 0.72 & 0.02 & 4.34 \\
		  & $\fcbm = 0.05$, $\alpha_{\rm{sc}} = 0.4$ & 11.02 & 7.50 & 8.83 & 6.64 & 9.97 & 0.63 & 0.01 & 4.28 \\
\hline
		$25 \rm{M}_\odot$, & $\fcbm = 0.004$ & 17.10 & 6.54 & 7.84 & 5.24 & 6.63 & 0.75 & 0.36 & 4.44 \\
		Schwarzschild, & $\fcbm = 0.01$ & 15.69 & 6.85 & 7.86 & 5.31 & 6.70 & 0.77 & 0.26 & 4.44 \\
		$f_0 = 0.002$ & $\fcbm = 0.022$ & 12.57 & 7.95 & 8.54 & 6.05 & 7.08 & 0.69 & 0.02 & 4.38 \\ 
		  & $\fcbm = 0.035$ & 14.03 & 9.00 & 9.73 & 7.24 & 7.43 & 0.64 & 0.02 & 4.34 \\
		  & $\fcbm = 0.05$ & 12.69 & 10.16 & 11.27 & 8.86 & 7.78 & 0.55 & 0.01 & 4.24 \\
\hline
		$25 \rm{M}_\odot$, & $\fcbm = 0.004$, $\alpha_{\rm{sc}} = 0.4$ & 21.35 & 6.43 & 7.77 & 5.18 & 6.53 & 0.71 & 0.72 & 4.45 \\
		Ledoux, & $\fcbm = 0.01$, $\alpha_{\rm{sc}} = 0.4$ & 21.59 & 6.91 & 7.72 & 5.15 & 6.70 & 0.70 & 0.78 & 4.43 \\
		$f_0 = 0.002$ & $\fcbm = 0.022$, $\alpha_{\rm{sc}} = 0.4$ & 14.87 & 7.94 & 8.59 & 6.11 & 7.08 & 0.66 & 0.12 & 4.38 \\
		  & $\fcbm = 0.035$, $\alpha_{\rm{sc}} = 0.4$ & 13.70 & 9.00 & 9.88 & 7.39 & 7.43 & 0.61 & 0.02 & 4.33 \\
		  & $\fcbm = 0.05$, $\alpha_{\rm{sc}} = 0.4$ & 12.76 & 10.16 & 11.68 & 9.25 & 7.78 & 0.52 & 0.04 & 4.24 \\
\bottomrule
	\end{tabular}
	\begin{tablenotes}
		\begin{footnotesize}
			\item \textbf{Notes:} Shown are the total star mass, $\rm{M}_{\rm tot}$, the helium core mass, $\rm{M}_\alpha$, the carbon-oxygen core mass, $\rm{M}_{\rm CO}$, the MS lifetime, $\tau_{\rm H}$, the core helium burning lifetime, $\tau_{\rm He}$ and the BSG to core helium burning lifetime, $\tau_{\rm BSG}/\tau{\rm He}$. The core mass is defined as the location where the abundance of the main fuel in the burning process, which creates the main end product of the burning phase, is below 0.1 and the abundance of the end product is above 0.01.
			\item[a] Hydrogen free core at hydrogen depletion.
			\item[b] $\tau_{\rm BSG}$, the BSG lifetime is defined as the time when the star (i) has left the MS stage of core hydrogen burning, (ii) the surface temperature is in the range $4.4 > \log_{10}\,\rm{T}_{\rm eff} > 3.9$ and (iii) it is not an extremely helium-enriched Wolf-Rayet-like star, i.e. $\rm{X}_{\rm surf} (^1 \rm{He}) > 0.3$.
			\item[c] The logarithm of the minimum effective temperature during the MS evolution. The terminal-age main-sequence is defined as the time when the central hydrogen mass fraction drops below $10^{-5}$.
		\end{footnotesize}
	\end{tablenotes}
		\label{properties_hedep}
	\end{threeparttable}
\end{table*}

\subsection{Nucleosynthesis in Core Helium Burning} \label{nucleosynthesis_heburn}

The two dominant reactions by which helium burns are the triple-$\alpha$ process, $3 \alpha\,\rightarrow\,^{12}$C, and $\alpha$ capture on carbon, $^{12}$C($\alpha,\gamma$)$^{16}$O. The first has a second order dependence on density whereas the latter a first order \citep[e.g.][]{Kippenhahn1994, Arnett1985, Woosley2002}. Therefore, with increasing abundance of $^{12}$C towards the end of core helium burning, the second reaction dominates. Furthermore, the density dependency favours the latter reaction at higher entropy, i.e. higher masses. Stars at solar metallicity additionally contain some $^{14}$N in their cores ($\sim 1.4\%$), which is left over from the CNO-cycle during the MS \citep{Arnould1993}. At the start of core helium burning, before the energy generation by the triple-$\alpha$ process becomes noteworthy, the $^{14}$N burns convectively away via $^{14}$N($\alpha,\gamma$)$^{18}$F($\beta^+\nu$)$^{18}$O($\alpha,\gamma$)$^{22}$Ne \citep{Cameron1960}. The synthesised $^{22}$Ne will capture another $\alpha$ nuclei once the central temperature exceeds $\sim 2.5 \times 10^8\,$K via $^{22}$Ne($\alpha$,n)$^{25}$Mg, creating the condition for the weak slow neutron capture process \citep[weak s-process;][]{Couch1974, Arnett1985, Prantzos1990, Kaeppeler1994, Frischknecht2016}. Yet only a part of the central $^{22}$Ne captures a $\alpha$ nuclei during core helium burning. The leftover $^{22}$Ne will capture a $\alpha$ during carbon shell burning, where the $\alpha$s are provided from the $\alpha$-emission channel of the $^{12}$C$+^{12}$C reaction. This creates the condition for the weak s-process at higher temperatures and slightly different conditions \citep[e.g.][]{Couch1974, Prantzos1990, Raiteri1991b, Pignatari2010}. This secondary neutron-source reaction competes during the late core helium burning with the $^{12}$C($\alpha,\gamma$)$^{16}$O reaction for the remaining $\alpha$ nuclei.\\
The outcome of core helium burning affects the further evolution of the star in several ways. The $^{12}$C to $^{16}$O ratio at core helium depletion depends strongly on the nucleosynthesis and its uncertainties \citep[e.g.][]{Arnett1985, Fields2018}. The outcome not only sets the fuel for the subsequent carbon and oxygen burning phases but also influences the pre-supernovae abundances \citep[e.g.][]{Thielemann1985, Woosley2002}. Furthermore, the amount of $^{12}$C available at core carbon burning ignition determines whether carbon burns convectively or radiatively, which has consequences on the convective history and the stellar structure at core collapse \citep{Ugliano2012, Sukhbold2014, Mueller2016, Sukhbold2016, Ertl2016, Sukhbold2018, Chieffi2019}. Moreover, the different activity of the $^{22}$Ne+$\alpha$ reaction during core helium or shell carbon burning will affect the nucleosynthesis and final weak s-process yields, because the burning conditions differ and there are different isotopic abundances, e.g. neutron poison, in the two stages \citep{Prantzos1990, Raiteri1991a, Raiteri1991b, Pignatari2010}.\\
CBM (i) increases the effective size of the convective helium core (Fig.~(\ref{convHeliumCoreMass})) and (ii) slightly changes the central conditions. Furthermore, the convective stability criterion influences the size of the convective core and when it grows. These facts affect the amount of $\alpha$ nuclei available during a certain period of core helium burning and the central conditions. This impacts the $^{12}$C to $^{16}$O ratio and the amount of $^{22}$Ne capturing an $\alpha$ nuclei, thus, the efficiency of the weak s-process during core helium burning \citep[see also e.g.][]{Costa2006}.\\
\begin{figure}
	\includegraphics[width=\columnwidth]{figures/legend_convHydrogenCore_15M_f00p002_colorblind_revision.png}
	\includegraphics[width=\columnwidth]{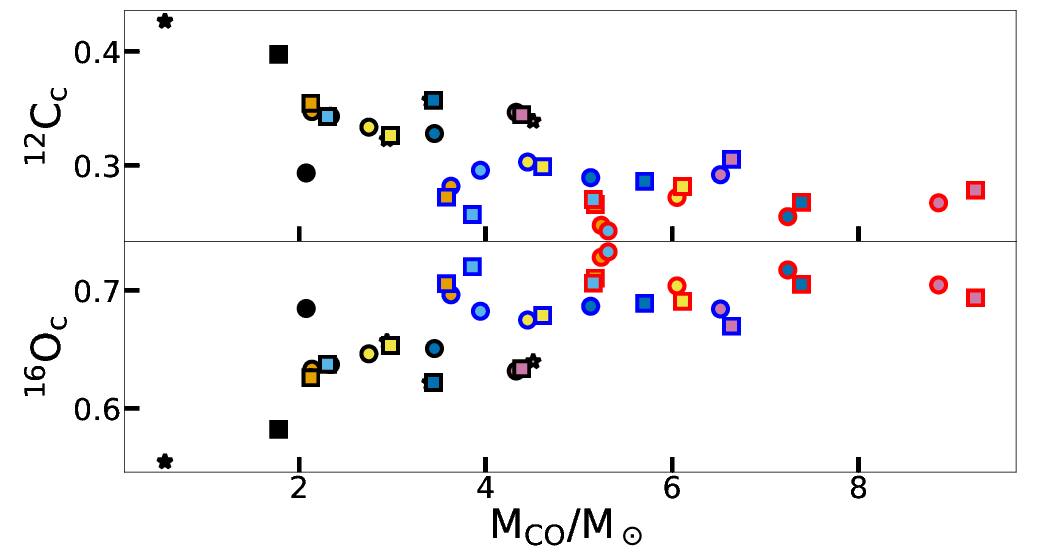}
	\caption{The $^{12}$C (top) and $^{16}$O (bottom) mass fractions in the centre as a function of the carbon-oxygen core mass at core helium depletion. Plotted are the values for the $15$ (black edge), $20$ (blue edge) and $25\,\Msun$ (red edge) models. The Schwarzschild boundary criterion is indicated by a circle and Ledoux criterion with a star ($\alphasc = 0.004$) or a square ($\alphasc = 0.4$), respectively. The color scheme for $\fcbm$ is the same as in Fig.~(\ref{boundaryLocation_hydrogenCore}).}
	\label{c12o16MassFraction_heliumDepletion}
\end{figure}
Fig.~(\ref{c12o16MassFraction_heliumDepletion}) presents the $^{12}$C and $^{16}$O mass fraction in the centre of the star at core helium depletion as a function of the carbon-oxygen core mass for the different initial masses. In the $15\,\Msun$ models the $^{12}$C mass fraction decreases and the $^{16}$O mass fraction increases with increasing $\fcbm$. The $15\,\Msun$ models with the two highest values of $\fcbm$ follow this trend but are shifted to slightly larger $^{12}$C and smaller $^{16}$O mass fractions. The first trend is a consequence of the increasing convective core mass during core helium burning with larger $\fcbm$ (Section \ref{ConvectiveHeliumCore}), which therefore ingests more $\alpha$-nuclei into the central burning zone. Thus, during late core helium burning, when the $^{12}$C$+\alpha$ is dominant, more $^{12}$C is turned into $^{16}$O. The second point is related to the timing of the convective helium core growth, see Fig.~(\ref{convHeliumCoreMass}). Whilst in the $15\,\Msun$ models with $\fcbm \leq 0.022$ the core growth is proportional, in the models with $\fcbm = 0.035$ or $0.05$ core growth occurs mainly during the initial phase of core helium burning (i.e. when X$_c$($^4$He) $\geq 0.8$) and the core growth is slower thereafter. Therefore, there is less $\alpha$ entrained during the late burning phase, when the $^{12}$C$+\alpha$ is dominant, resulting in a slightly higher $^{12}$C and lower $^{16}$O mass fraction in these models. The Schwarzschild model with $\fcbm = 0.035$ has a convective core growth more similar to the models with lower amounts of CBM. This is also reflected in its lower $^{12}$C and higher $^{16}$O compared to the rest of the models with large amounts of CBM.\\
Fig.~(\ref{c12o16MassFraction_heliumDepletion}) also depicts that within the semiconvective efficiencies used in this work, semiconvection has almost no influence on the carbon and oxygen mass fraction at the end of core helium burning. The only models where there is an impact are the ones with no CBM, where a higher semiconvective efficiency leads to a larger convective core (Fig.~(\ref{convHeliumCoreMass})). Consequently, there is more fuel available during the late burning phase, resulting in the lower $^{12}$C and higher $^{16}$O mass fraction.\\
The $15\,\Msun$ Schwarzschild model with no CBM does not follow this trend. This model has a much higher $^{16}$O to $^{12}$C ratio in the centre than any other of the $15\,\Msun$ simulations. This is because of the sharp increase of the convective core described in Section \ref{ConvectiveHeliumCore}, which transports fuel into the convective region during late core helium burning. This brings a lot of fresh $\alpha$ into the core and more carbon is synthesised into oxygen.\\
Fig.~(\ref{c12o16MassFraction_heliumDepletion}) clearly shows a higher $^{16}$O to $^{12}$C ratio with increasing initial mass. This is expected due to the temperature dependence of the $^{12}$C+$\alpha$ reaction, hence, more massive stars synthesise more $^{16}$O and less $^{12}$C during core helium burning \citep[e.g.][]{Prantzos1990}.\\
The central $^{12}$C and $^{16}$O mass fractions in the $20$ and $25\,\Msun$ models with different amounts of CMB appear to be constant around X$_c$($^{12}$C)$ \sim 0.3$, X$_c$($^{16}$O)$ \sim 0.68$ and X$_c$($^{12}$C)$ \sim 0.22$, X$_c$($^{16}$O)$ \sim 0.7$, respectively. Contrary to the $15\,\Msun$ models the $^{12}$C+$\alpha$ reaction seems to be saturated under the central conditions in these models.This is a result of the higher central temperatures in the models with the same initial mass but larger amounts of CBM, which leads other reactions to activate, such as $^{16}$O($\alpha,\gamma$)$^{20}$Ne \citep[e.g.][]{Arnett1985}. The models with values of $\fcbm \leq 0.01$ have a lower $^{12}$C and higher $^{16}$O mass fraction. This behaviour is a consequence of the \textit{core breathing pulses}, which occur during the late stages of core helium burning and affect the size of the convective core (see discussions in Sections \ref{ConvectiveHeliumCore} and \ref{discussion}). These events ingest more fuel into the burning region. Therefore the final $^{16}$O mass fraction is increased and the $^{12}$C mass fraction is decreased. Since the \textit{core breathing pulses} are more extreme with less CBM, the $^{16}$O to $^{12}$C ratio is slightly higher with less CBM.\\
Comparing the different initial masses it is obvious that depending on the amount of CBM lower mass models can have a similar $^{16}$O to $^{12}$C ratio at the end of core helium burning to higher mass models with less CBM. This is striking because this ratio is crucial in determining the evolution of the advanced burning phases, in particular the convective history \citep[see also discussions in e.g.][]{Sukhbold2014, Chieffi2019}. Therefore, the amount of CBM not only directly affects these convective regions by enhancing the convectively mixed region but also indirectly by setting the $^{16}$O to $^{12}$C ratio at the end of core helium burning.\\
The carbon to oxygen ratio at core helium depletion is very uncertain, especially due to the uncertainty in the reaction rate \citep[e.g.][]{Woosley2002}. We show here, that the amount of CBM introduces another uncertainty in this ratio. The convective boundary criterion seem to affect the ratio less (Fig.~(\ref{c12o16MassFraction_heliumDepletion})). Nevertheless, the importance of this uncertainty needs to be determined in the absence of \textit{core breathing pulses} which are thought to be numerical artefacts \citep[][see discussion in Section \ref{discussion}]{Constantino2016} but see Section \ref{discussion}.\\
\begin{figure}
	\includegraphics[width=\columnwidth]{figures/legend_convHydrogenCore_15M_f00p002_colorblind_revision.png}
	\includegraphics[width=\columnwidth]{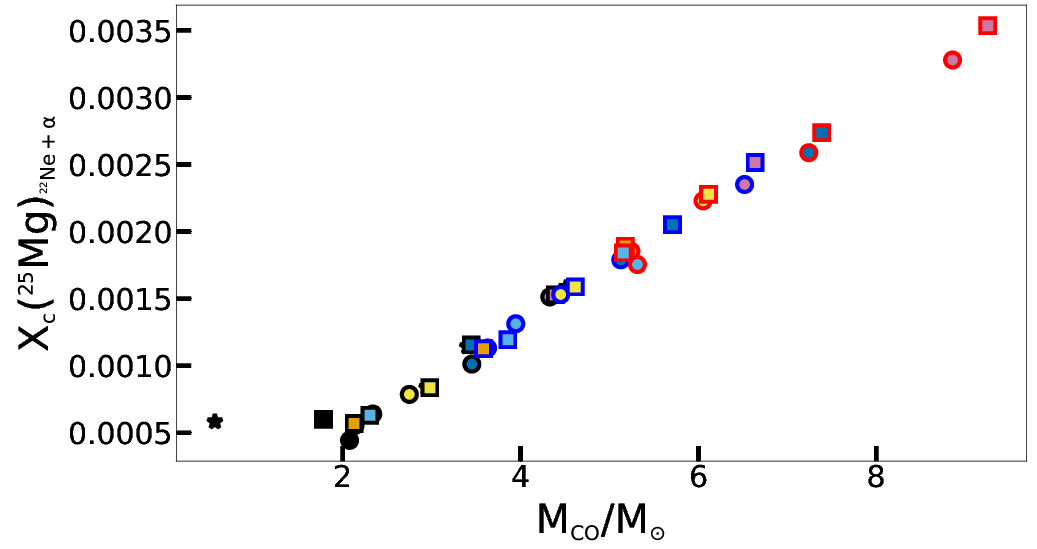}
	\caption{The $^{25}$Mg mass fraction produced at the centre by the neutron source reaction during core helium burning as a function of the carbon-oxygen core mass at core helium depletion. Plotted are the values for the $15$ (black edge), $20$ (blue edge) and $25\,\Msun$ (red edge) models. A circle indicates the \textit{Schwarzschild} criterion, a star the \textit{Ledoux} criterion with $\alphasc = 0.004$ and a square the \textit{Ledoux} criterion with $\alphasc = 0.4$. The color scheme is the same as in Fig.~(\ref{boundaryLocation_hydrogenCore}).}
	\label{ne22mg25MassFraction_heliumBurn}
\end{figure}
The weak s-process in massive stars depends on the efficiency of the neutron source reaction $^{22}$Ne($\alpha$,n)$^{25}$Mg, which determines the neutron density in the helium core. The efficiency of the neutron source reaction depends on (i) the nuclear abundances of $^{22}$Ne and $^4$He and (ii) the central conditions such as temperature and density. At solar metallicity, most of the metals are CNO elements, which are mostly turned into $^{14}$N during hydrogen burning. Hence, the $^{22}$Ne abundance before the activation of the $^{22}$Ne+$\alpha$ reaction during core helium burning is directly related to the initial metal abundance in all our calculations (X($^{22}$Ne) $\approx\frac{22}{14}\cdot$X$^{\rm CNO}_{ini}$). The amount of $\alpha$ available, on the other hand, depends on the size of the convective core and on the amount of fuel entrained at the top of the convective core. The differences in the central thermodynamic condition between the models depend on the different amounts of CBM as well.\\
Fig.~(\ref{ne22mg25MassFraction_heliumBurn}) presents the $^{25}$Mg mass fraction  produced at the centre by the neutron source reaction $^{22}$Ne($\alpha$,n)$^{25}$Mg during core helium burning as a function of the carbon-oxygen core mass for the three initial masses. The amount of neutrons released by the neutron source reaction is equal to the number of $^{25}$Mg produced in Fig.~(\ref{ne22mg25MassFraction_heliumBurn}), thus, it indicates the neutron density and with it the activity of the weak s-process.\\
Fig.~(\ref{ne22mg25MassFraction_heliumBurn}) shows a clear trend of an increasing $^{25}$Mg production with increasing $\fcbm$ for all initial masses. This is because (i) more fuel is entrained into the convective zone with an increasing amount of CBM and (ii) the models with more CBM have larger core masses, hence they burn helium in the centre at a slightly higher temperature and lower density compared to models with less CBM. Moreover, the convective boundary criterion leads to different convective core sizes (Fig.~(\ref{convHeliumCoreMass})), which leads to different activity of the neutron source reaction. The shift of $\rm{X}_c(^{25}\rm{Mg})$ to higher values with larger initial mass results from the fact that these models burn helium at higher temperatures.\\
Fig.~(\ref{ne22mg25MassFraction_heliumBurn}) depicts that the different semiconvective efficiencies do not change the amount of $^{25}$Mg produced by the $^{22}$Ne+$\alpha$ reaction, which is a result of the reduced occurrence of semiconvection with increasing amount of CBM. Therefore, the semiconvective efficiency does not affect the weak s-process efficiency during core helium burning.\\
Again, it is interesting to see in Fig.~(\ref{ne22mg25MassFraction_heliumBurn}) that models with a large amount of CBM behave like models of a larger initial mass but less CBM in terms of cabon-oxygen core mass and s-process activity.\\
We want to stress here, that the $\rm{X}_c(^{25}\rm{Mg})$ in Fig.~(\ref{ne22mg25MassFraction_heliumBurn}) is not equal to the $^{25}$Mg abundance at core helium depletion. Indeed, $\rm{X}_c(^{25}\rm{Mg})$ corresponds to the production at the very centre, whereas the final $^{25}$Mg abundance is determined by the conditions throughout the convective core. Additionally, some of the $^{25}$Mg is further processed by burning.\\
The difference in s-process activity during core helium burning can shift the final weak s-process production between iron and strontium. Furthermore, more $^{22}$Ne$+\alpha$ consumption during core helium burning leads to higher s-process yields \citep{Pignatari2010}.\\
The changes in nucleosynthesis due to the $f_0$-parameter is linear. A larger $f_0$ implies slightly less CBM, thus, lower amount of fuel available in the late core helium burning. Therefore, the abundances of $^{12}$C and $^{22}$Ne are larger and the abundances of $^{16}$O and $^{25}$Mg are lower than shown in Figs.~(\ref{c12o16MassFraction_heliumDepletion}) and (\ref{ne22mg25MassFraction_heliumBurn}). The differences of the nucleosynthesis during core helium burning due to semiconvection are slim as can be seen in Figs.~(\ref{c12o16MassFraction_heliumDepletion}) and (\ref{ne22mg25MassFraction_heliumBurn}) and mainly affect the simulations with no CBM.

\section{Blue versus Red Super-Giant Evolution}\label{BSG_vs_RSG}

The different behaviour in depth and duration of the ICZ discussed in Section \ref{ICZ} has an important impact on the surface evolution of the model, which in turn influences the later stellar structure and evolution.\\
\begin{figure}
	\includegraphics[width=\columnwidth]{figures/legend_convHydrogenCore_15M_f00p002_colorblind_revision.png}
	\includegraphics[width=\columnwidth]{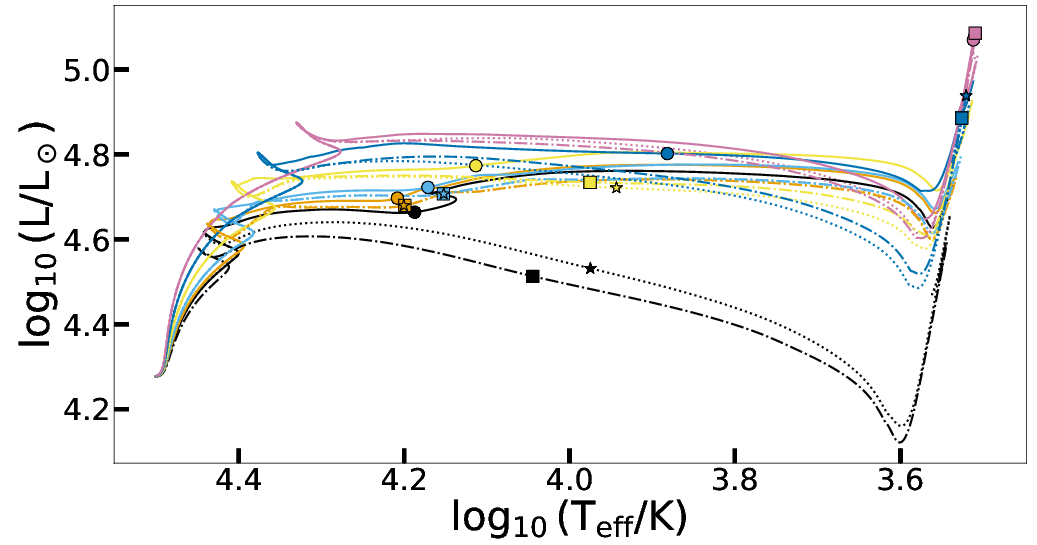}\\
	\includegraphics[width=\columnwidth]{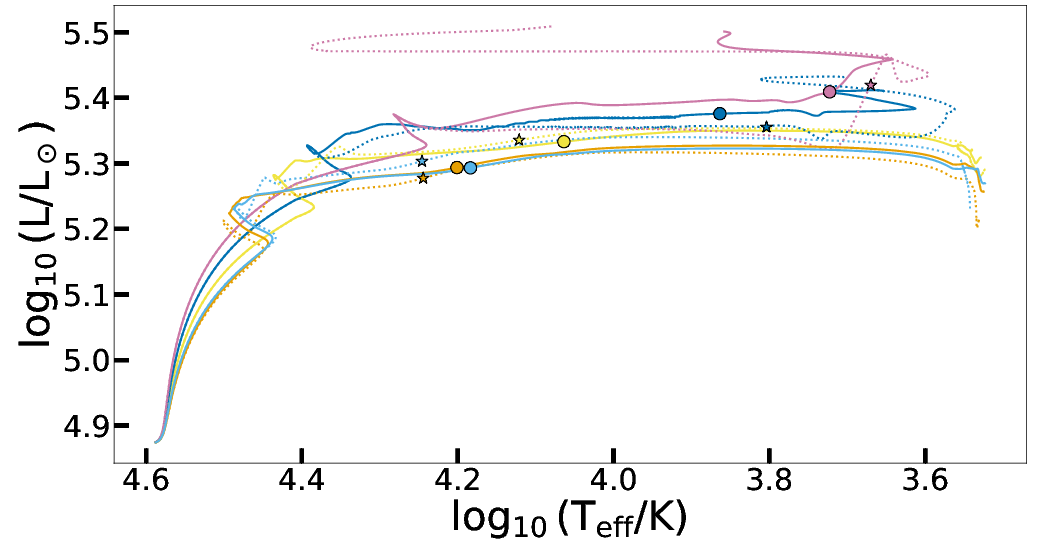}
	\caption{The HRD for various calculations of models with initial masses of $15$ (top) and $25\,\Msun$ (bottom), all with $f_0 = 0.002$. The solid lines indicate Schwarzschild models and all the others are Ledoux models with either $\alpha_{\rm{sc}} = 0.4$ (dotted line) or $\alpha_{\rm{sc}} = 0.004$ (dash-dotted line). The color scheme indicates the amount of CBM. The markers show the location where helium burning is ignited in the core (0.3\% of the helium left after core hydrogen depletion is burnt). The different marker styles indicate the different boundary criteria used in the calculation, where circles indicate the \textit{Schwarzschild} criterion and the others are Ledoux models with $\alpha_{\rm{sc}} = 0.004$ (square) or $\alpha_{\rm{sc}} = 0.4$ (star).}
	\label{HRD}
\end{figure}
Fig.~(\ref{HRD}), top panel, shows the HRD for the $15\,\Msun$ models. We discussed the MS-tracks in the HRD in Section \ref{HydrogenBurning}. The models leave the MS via the \textit{Henyey hook} where the first difference between the two boundary criteria arises. The tracks of the Ledoux models form a loop before they start crossing towards the cooler, red side of the HRD, whereas the Schwarzschild models evolve via a hook. With increasing amount of CBM, the latter show loops as well. This contrast arises from the different location of the intermediate convective zone, which becomes more similar with increasing $\fcbm$ (see Figs.~(\ref{kippenhahn_15M}) and (\ref{kippenhahn2_15M})).\\
The Ledoux models with no CBM (black dotted and dashdotted lines) show the most different tracks when crossing to the RSG branch on the cool side of the HRD by decreasing their surface luminosity.
\begin{figure}
	\includegraphics[width=\columnwidth]{figures/legend_convHydrogenCore_15M_f00p002_colorblind_revision.png}
	\includegraphics[width=\columnwidth]{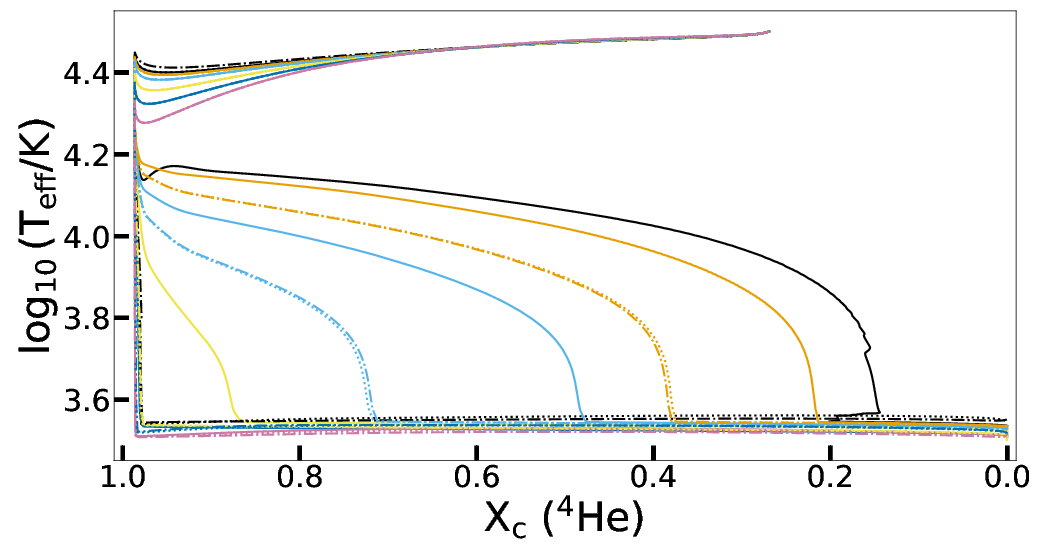}
	\includegraphics[width=\columnwidth]{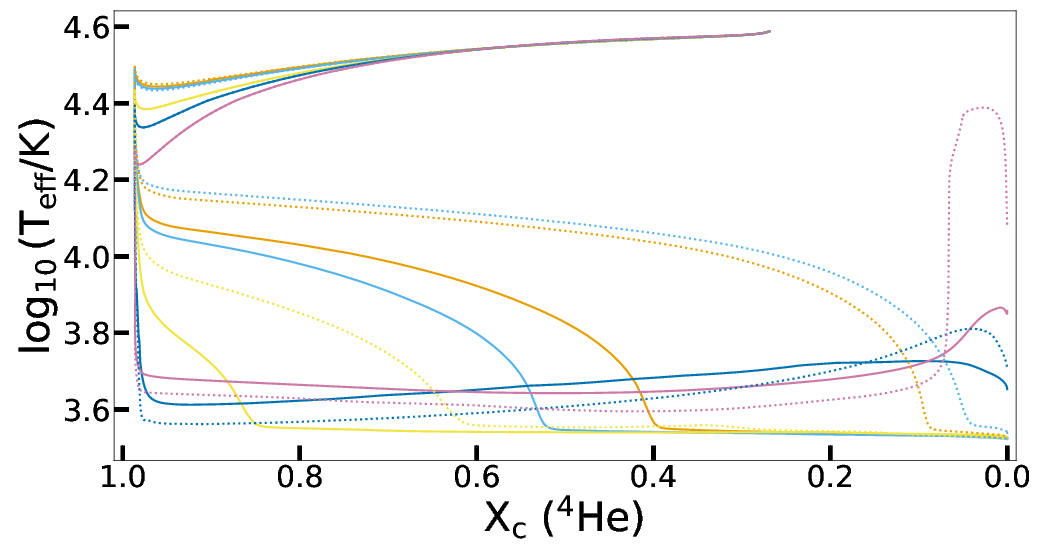}
	\caption{Evolution of the effective temperature as a function of the central $^4$He mass fraction. Shown are the $15$ (top) and $25\,M_\odot$ (bottom) models, all with $f_0 = 0.002$. The solid lines indicate Schwarzschild models and all the other are Ledoux models with either $\alpha_{\rm{sc}} = 0.4$ (dotted line) or $\alpha_{\rm{sc}} = 0.004$ (dash-dotted line). The color scheme indicates the amount of CBM.}
	\label{Teff_XcHe4}
\end{figure}
Fig.~(\ref{Teff_XcHe4}), top panel, which presents the evolution of the surface temperature as a function of the central helium mass fraction sheds light on what happens in the interior of these models. The two Ledoux models, after the \textit{Henyey hook} (upper left corner), drop their surface temperature before they consume any significant amount of helium in their interior. Hence, they directly cross the \textit{Herzsprung gap} to the cool side of the HRD before they fully ignite helium in their core, despite the fact that Fig.~(\ref{HRD}) suggests that they start burning helium at $\log T_{\rm{eff}} \approx 4.0$. The fast red-ward evolution results in a primarily adiabatic expansion of the envelope, thus the decrease in surface luminosity. The Schwarzschild model with no CBM (black solid line), on the other hand, ignites helium in its core at around $\log T_{\rm{eff}} \approx 4.2$ in Fig.~(\ref{HRD}). Fig.~(\ref{Teff_XcHe4}) reveals, that the model consumes about $\sim 80\%$ of its central $^{4}$He in the hotter, blue side of the HRD before it moves to the cool, red side. The crossing of the \textit{Herzsprung gap} in this simulation occurs with a nearly constant surface luminosity, which is due to a quasi-hydrostatic contraction/expansion. The short loop-like structure at $\log T_{\rm{eff}} \approx 4.2$ in the Schwarzschild model is because of a strong, boosted hydrogen shell, which stops the expansion of the envelope by supporting the core.\\
These two scenarios, either spending the whole of core helium burning phase as a RSG or spending most of core helium burning as a blue super-giant (BSG), are a consequence of the different ICZ and the activity of the hydrogen shell (see Section \ref{ICZ}).\\
These extreme differences of red- or blue-ward evolution with either the \textit{Ledoux} or the \textit{Schwarzschild} criterion in the $15\,\Msun$ models decrease with increasing amount of CBM (Figs.~(\ref{HRD}) and (\ref{Teff_XcHe4})). However, crucial variations arise with the choice of $\fcbm$. When a small amount of CBM is applied ($\fcbm = 0.004$), the erasing of the $\mu$-gradient  at the lower convective boundary of the ICZ in the Ledoux models dominates (see Section \ref{ICZ}). As a result, both, the Ledoux and the Schwarzschild models, predict an overlap of the ICZ and the hydrogen burning shell. Consequently, these models experience a more `\textit{Schwarzschild}-like' evolution where the star consumes about $\sim 50\%$ of its central helium before they evolve red-wards. The BSG lifetimes is shorter in these models compared to the Schwarzschild model with no CBM (Table \ref{properties_hedep}) because of the shorter duration of the ICZ (Fig.~(\ref{luminosities_15M})). These three models exhibit a vertex at $\log T_{\rm{eff}} \approx 4.2$ which is due to the boost of the hydrogen shell. When higher amounts of CBM are used ($\fcbm = 0.01$, $0.022$ and $0.035$ in the Schwarzschild case), the effect of faster energy regulation dominates (see Section \ref{ICZ}). This results in a shorter duration of the ICZ and a weaker hydrogen shell. As a consequence, the models evolve more `\textit{Ledoux}-like', meaning they evolve red-wards faster and spend most of their core helium burning phase as RSGs. The loop-like structure around $\log T_{\rm{eff}} \approx 4.2$ in Fig.~(\ref{HRD}) flattens and the point of central helium ignition is shifted to slightly cooler values of the surface temperature. For even more CBM ($\fcbm = 0.05$ and $0.035$ for the Ledoux models) the ICZ does not overlap with the hydrogen burning shell anymore (even less than in the Ledoux models with no CBM). Therefore, the energy generation of hydrogen burning decreases fast and the shell moves outwards due to the lack of fuel at the shell location (see Section \ref{ICZ}). Consequently, the model moves very fast to the cooler part of the HRD (Fig.~(\ref{Teff_XcHe4}) and Table \ref{properties_hedep}). Moreover, core helium burning only ignites once the model has ascended the red giant branch. This is in fact the result from an even faster red-ward evolution than the other models. The Ledoux models with no CBM and with $\fcbm = 0.022$ and the Schwarzschild model with $\fcbm = 0.035$ reach the RSG branch after they have consumed $\sim 0.8\%$ of their X$_c(^4\rm{He})$. On the contrary, the Ledoux models with $\fcbm = 0.035$ reach the RSG branch with having less helium consumed and the models with $\fcbm = 0.05$ do not consume any notable amount of helium before they start ascending the RSG branch. Since these models do not have a strong hydrogen shell to support the core against contraction, more gravitational energy is released, of which part leads to a more extreme expansion and relatively high mass-loss rates before helium is ignited in the centre. Hence the point of core helium ignition is after the models starts moving up the RSG branch. The effect of this transition, between `\textit{Schwarzschild}-like' to `\textit{Ledoux}-like', is nicely presented in Fig.~(\ref{Teff_XcHe4}) and is a function of $\fcbm$.\\
The generally larger ICZ in the $20$ and $25\,\Msun$ models (see Section \ref{ICZ}) leads to a slower red-wards evolution and these models generally consume more helium in their cores as BSGs (Figs.~(\ref{HRD}) and (\ref{Teff_XcHe4}), both bottom panel). The location of helium ignition in Fig.~(\ref{HRD}), however, does not greatly change with higher initial mass and seems to be more dependent on the amount of CBM.\\
Interestingly, the central evolution presented in Fig.~(\ref{Teff_XcHe4}) reveals important qualitative differences. The Ledoux models with $\fcbm = 0.004$ and $0.01$ both spend nearly their whole core helium burning phase as BSGs ($\sim 90\%$ and $\sim 75\%$, respectively, see Table \ref{properties_hedep}), whereas their Schwarzschild counterparts only spend about $\sim 64\%$ and $\sim 30\%$, respectively, of their core helium burning phase as BSGs. Furthermore, the Ledoux models with $\fcbm = 0.004$ and $0.01$ behave more similarly during this stage than the respective Schwarzschild model with the same $\fcbm$ and vice-versa. Hence, the stability criterion introduces a larger uncertainty in these models with medium and low amounts of CBM during this stage. This occurs for both initial masses, $20$ and $25\,\Msun$.\\
Another interesting feature appearing in Figs.~(\ref{HRD}) and (\ref{Teff_XcHe4}) is that the models with large amount of CBM ($\fcbm = 0.035$ and $0.05$) start to move back to the hot side of the HRD. This is a consequence of the fast red-wards evolution after the MS of these models and the resulting large mass-loss (see Fig.~(\ref{MassLossRates_XcHe4_15M_f00p002_variousMixing})), which erodes most of the hydrogen rich envelopes of these stars (compare $\rm{M}_{\rm tot}$ and $\rm{M}_\alpha$ in Table \ref{properties_hedep}). Indeed, the $25\,\Msun$ models with $\fcbm = 0.05$ lose enough mass to evolve blue-wards with the Ledoux model ending in the WR phase (X$_{\rm surf}$($^1$H) $< 0.4$ and log$_{10}(T_{\rm eff}>4.0$). As expected, this behaviour is more dominant in the $25\,\Msun$ models than in the $20\,\Msun$ models.\\
The effect of the \textit{convective fingers} on the ICZ in the models with $\fcbm = 0.022$, as discussed in Section \ref{HydrogenBurning}, is also apparent. In the $20\,\Msun$ Ledoux model the duration of the ICZ is shorter due to the influence of the \textit{convective fingers} and it crosses nearly directly to the cool part of the HRD, whereas its Schwarzschild counterpart stays for some time in the hot part. In the $25\,\Msun$ Ledoux models with $\fcbm = 0.022$, on the other hand, the ICZ is larger due to the interference of the \textit{convective fingers} and it experiences a slower red-wards evolution than its Schwarzschild counterpart.\\
Table \ref{properties_hedep} includes the ratio of the BSG to core helium burning lifetimes. This shows once more that the simulation with a larger and longer ICZ stay for longer in the hot part of the HRD. Vice-versa, a smaller ICZ leads to a faster red-wards evolution. Furthermore, the Schwarzschild models predict constant or slightly decreasing BSG lifetimes with increasing initial mass whereas the Ledoux models predict first an increase and then a decrease for the two largest values of $\fcbm$. This difference is due to the initial location of the ICZ.\\
Some of the models spend more than half of their core helium burning lifetime as BSGs. This is partly in contradiction with \citet{Davies2018} who found a transition from fast to slow red-ward evolution around $15 - 20 M_\odot$. Moreover, \citet{Davies2018} state that the location of helium ignition impacts the way the star crosses the HRD. We, however, do not see a clear indication of this in our mass range because all of the models with $\fcbm \lesssim 0.022$ ignite helium burning in their cores at around log$_{10}(T_{\rm eff} \approx 4.2-4.1$ with the exceptions being the models with large amount of CBM. So the location of helium ignition in our models only shows whether a low and intermediate $\fcbm$ or a large $\fcbm$ is used. The main impact we find, as stated by \citet{Davies2018} as well, is the location and duration of the ICZ.\\
These two different evolutionary paths, core helium burning as a BSG or RSG (or a combination thereof), have an important impact on the structure of the star and its further evolution, in particular (i) the mass loss rates, (ii) the shape of the surface convective zone and surface enrichment, (iii) the central evolution (Section \ref{heliumBurning}) and (iv) the star type at the end of core helium burning (RSG, BSG or WR star).\\
\begin{figure}
	\includegraphics[width=\columnwidth]{figures/legend_convHydrogenCore_15M_f00p002_colorblind_revision.png}
	\includegraphics[width=\columnwidth]{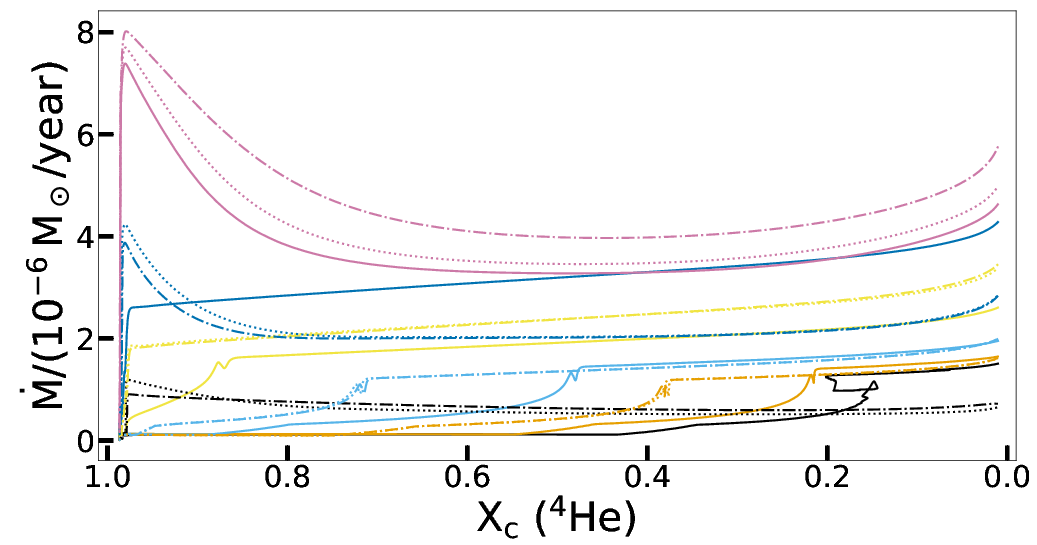}
	\includegraphics[width=\columnwidth]{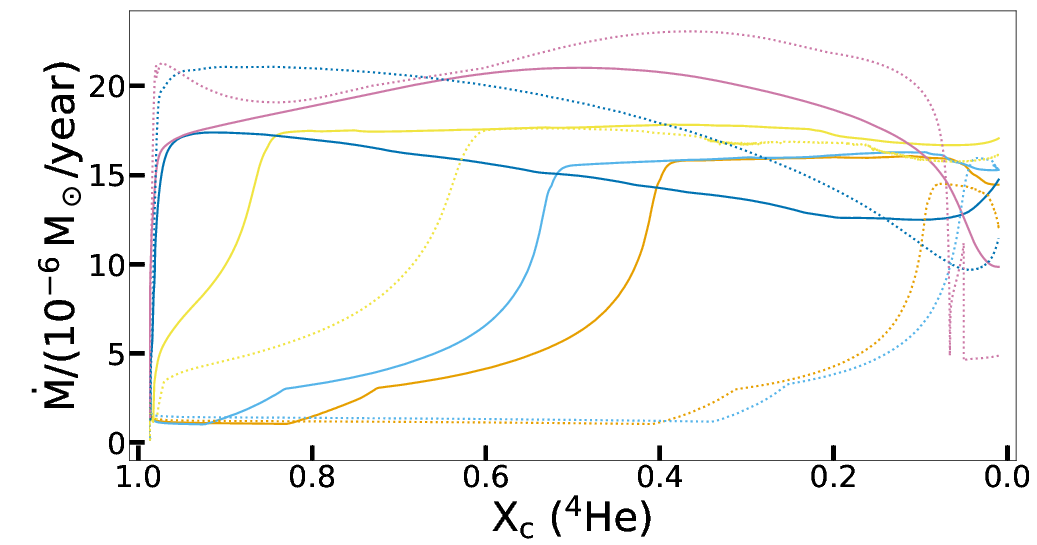}
	\caption{The mass loss rate per year as a function of the central helium mass fraction. The tracks begin at central hydrogen depletion and go up to central helium depletion. Shown are the tracks for the $15$ (top) and $25\,\Msun$ (bottom) models, all with $f_0 = 0.002$. The solid lines indicate Schwarzschild models and all the other lines are Ledoux models with either $\alpha_{\rm{sc}} = 0.4$ (dotted line) or $\alpha_{\rm{sc}} = 0.004$ (dash-dotted line). The color scheme is shown by the color scheme.}
	\label{MassLossRates_XcHe4_15M_f00p002_variousMixing}
\end{figure}
Fig.~(\ref{MassLossRates_XcHe4_15M_f00p002_variousMixing}) presents the mass loss rates, $\dot{\rm M}$, per year as a function of the central helium mass fraction, starting at core hydrogen depletion (bottom left). $\dot{\rm M}$ shows similar low values in all models at the end of their MS evolution. Thereafter the mass loss rates evolution exhibits quite a contrast. The mass loss rates depend crucially on the time when the model enters the RSG phase. During the time a model evolves as a BSG, its mass loss rates stay relatively low, following the mass loss prescription from \citet{Vink2000, Vink2001}. However, once the model evolves red-wards it experiences a drastic increase in its mass loss rate. This is around log$_{10}(T_{\rm eff}=4.0$ when the stellar evolution code switches from the mass loss prescription for O-stars to the empirical mass loss rates by \citet{deJager1988}. The different dependency on the surface temperature and luminosity lead to the drastic increase of the mass loss rates.\\
As previously discussed, the time a model spends as a BSG depends on the duration, location and size of the ICZ, which itself depends on the stability criterion used and the amount of CBM applied. In general, models with a larger $\fcbm$ parameter experience a faster red-wards evolution, hence, they increase $\dot{\rm M}$ earlier. Additionally, there is a difference between the \texttt{Ledoux} and the \texttt{Schwarzschild} criterion models, which is present for all $\fcbm$-values used in this work. This is a result from the differences of the ICZ discussed in Section \ref{ICZ}.\\
The simulations with no CBM (black lines in Fig.~(\ref{MassLossRates_XcHe4_15M_f00p002_variousMixing}), top panel) fall out of this general trend, which is because the Schwarzschild model has the longest BSG lifetime whereas the Ledoux models the shortest. Furthermore, the models with $\fcbm = 0.05$ and the Ledoux models with $\fcbm = 0.035$ experience a relatively high $\dot{M}$ which is reduced by about half in the subsequent evolution. This is due to the drop of the total luminosity (Fig.~(\ref{luminosities_15M})). The high luminosity after core hydrogen depletion in these models is due to the more extreme central contraction without a strong support from the hydrogen shell (Section \ref{ICZ}).\\
The $20$ and $25\,\Msun$ models generally have higher mass-loss rates with increasing $\fcbm$ and hence a smaller total mass at core helium depletion (Table \ref{properties_hedep}). There are, however, some exceptions. The Ledoux and Schwarzschild models, each with $\fcbm = 0.004$ and $0.01$, show more similar mass loss rates than their counterpart with the same $\fcbm$ but different stability criterion. This is a consequence of the more similar ICZ previously discussed.\\
The changes in mass-loss rates seen in Fig.~(\ref{MassLossRates_XcHe4_15M_f00p002_variousMixing}) are due to the different phases the models are in, as discussed above. The variation due to the uncertainty of CBM leads to a wide range of total masses at core helium depletion (Table \ref{relativeVariation_coremasses}). The most extreme case are the models with an initial mass of $25\,\Msun$ where the final mass ranges from $21.59\,\Msun$ down to $12.57\,\Msun$.\\
The depth and strength of the ICZ further affects the appearance of the surface convective zone. A less energetic hydrogen shell favours core contraction, thus expansion of the envelope, which in turn cools down and the opacity in the envelope increases. As a consequence, a surface convective zone develops, which penetrates deep into the star (see Figs.~(\ref{kippenhahn_15M}) and (\ref{kippenhahn2_15M}), where some panels show a convective zone in the upper right side) and enriches the surface with previously synthesised material from the interior. A stronger hydrogen shell, on the other hand, delays the formation of the surface convective zone and the surface enrichment occurs later \citep[see discussion in e.g.][]{Georgy2014}. Therefore, the surface composition and composition of winds will be affected.\\
In Section \ref{ICZ} we have shown that the $f_0$ parameter has an important effect on the ICZ when small values of $\fcbm$ are used, especially in the Ledoux models. According to our discussion above, this affects the red-wards evolution (Table \ref{properties_hedep}) and with it the mass loss rates. Therefore, this parameter should not be overlooked in discussions.

\section{Discussion}\label{discussion}

\begin{table*}
	\centering
	\begin{threeparttable}
	\caption{The absolute and relative variation of the total mass, $\rm{M}_{\rm tot}$, the helium core mass, $\rm{M}_\alpha$, and the carbon-oxygen core mass, $\rm{M}_{\rm CO}$. The individual values of each model are shown in Table \ref{properties_hedep}.}
	\begin{tabular}{ccccccc}\\
\toprule
			  & $\Delta\rm{M}_{\rm tot}/\rm{M}_\odot$ & $\delta\rm{M}_{\rm tot}$ & $\Delta\rm{M}_\alpha/\rm{M}_\odot$ & $\delta\rm{M}_\alpha$ & $\Delta\rm{M}_{\rm CO}/\rm{M}_\odot$ & $\delta\rm{M}_{\rm CO}$ \\
			\midrule
			$15\,\Msun$ & 3.38 \textcolor{blue}{(3.45)} & 27.72\% \textcolor{blue}{(27.80\%)} & 2.30 \textcolor{blue}{(2.74)} & 43.07\% \textcolor{blue}{(51.31\%)} & 2.40 \textcolor{blue}{(3.94)} & 69.77\% \textcolor{blue}{(114.53\%)} \\[0.5em]
			$20\,\Msun$ & 8.60 & 74.35\% & 2.94 & 36.70\% & 2.94 & 51.49\% \\[0.5em]
			$25\,\Msun$ & 9.02 & 65.84\% & 3.96 & 40.08\% & 4.1 & 55.48\% \\
\bottomrule
	\end{tabular}
		\label{relativeVariation_coremasses}
		\begin{tablenotes}
		\begin{footnotesize}
			\item \textbf{Notes:} The variation for a quantity $Q$ is calculated as 
			\begin{equation}
				\delta Q = \frac{\Delta Q}{Q_{\rm ref}} \equiv \frac{Q_{\rm max} - Q_{\rm min}}{Q_{\rm ref}} \times 100,
			\end{equation}
			where $Q_{\rm max}$ and $Q_{\rm min}$ are the maximal and minimal value of the quantity for the initial mass and $Q_{\rm ref}$ the value of the reference model (see text).
			\item[*] The blue values in brackets include the models with no CBM.
		\end{footnotesize}
	\end{tablenotes}
	\end{threeparttable}
\end{table*}
One of the main goals of this work is to show the relative importance of CBM uncertainties and which quantities of stellar evolution are mostly affected. In Table \ref{relativeVariation_coremasses} we list the variation of the core masses and the total mass at core helium depletion and in Table \ref{relativeVariation_lifetimes} the variation of the lifetime of some stellar stages. The two tables show for each initial mass the maximal difference of the predicted values of our simulations and the relative variation with respect to a reference model. The reference model for all initial masses is the Ledoux model with $\fcbm = 0.035$ and $\alphasc = 0.4$ , which is, in the $15\,\Msun$ case, the closest setting to the calibration of \citet{Brott2011}.\\
The two largest relative deviations in Table \ref{relativeVariation_coremasses} are the total mass of the star and the carbon-oxygen core mass, which are both above $50\%$. In the two higher initial masses, the uncertainty of the total star mass dominates ($\sim 65\%$ and $\sim 75\%$, respectively), whereas in the $15\,\Msun$ models the relative variation of the carbon-oxygen core is the largest with $\sim 70\%$. The absolute deviation of the total mass is about $9\,\Msun$. This large uncertainty is a result of the models spending different amounts of times as BSG or RSG, where different mass loss prescriptions apply. This is a consequence of both, the boundary criterion and the amount of CBM which influence the location, shape and duration of the ICZ, see Section \ref{ICZ}. The core masses, with the exception of the carbon-oxygen core mass in the $15\,\Msun$ models, show smaller but still non-negligible deviations. The helium core masses have differences up to $4\,\Msun$ and a relative variations of $\sim 35 - 45 \%$. The absolute difference is larger for the higher initial masses, a consequence of the size difference of these models, but the relative variation is smaller because of the larger $\Mhe$ of the reference model. The uncertainty of helium core mass is dominated by the amount of CBM and the choice of the boundary criterion only gives maximal differences up to $0.5\,\Msun$ (Table \ref{properties_hedep}). The carbon oxygen-core masses follow the same trend but with larger absolute and relative variations ($\sim 2.5 - 4.1\,\Msun$ and $\sim 50 - 70 \%$, respectively). These differences are mainly influenced by the choice of $\fcbm$ and the choice of the boundary criterion is less important than for $\Mhe$ in most cases. The boundary criterion mainly influences the timing when the convective helium core grows but has less impact on its maximal extent. $\Mco$ shows a slightly higher absolute variation because the variations cumulate, thus, the relative uncertainty of the core masses increases as stellar evolution proceeds and might be even higher for the further evolution \citep[see e.g.][]{Davis2019}. The uncertainty of the mixing assumptions also influences convective shell interactions during the later evolutionary stages \citep[e.g.][]{Clarkson2020}.\\
We want to stress here, that some of the core masses in the $15\,\Msun$ models with larger values of $\fcbm$ are as large as the same core mass in the $20\,\Msun$ models with moderate values of $\fcbm$. The same applies for the $20$ and $25\,\Msun$ models. These models with large amounts of CBM would therefore have an evolution after core helium burning that is more similar to models with a higher mass but less CBM. This would change the ZAMS - SN progenitor relation and the final fate of massive stars. Furthermore, the core masses are often used to relate to the pre-supernovae compactness and explodability of a star \citep[e.g.][]{OConnor2011, Mueller2016, Ertl2016, Sukhbold2018, Chieffi2019}. Relative uncertainties of $\gtrsim 40-70 \%$ make these predictions unreliable and more dependent on the parameter choices than the actual physics. Moreover, these uncertainties will impact 3D hydrodynamics simulations for which 1D stellar evolution models are used as input model.\\
\begin{table*}
	\centering
	\begin{threeparttable}
	\caption{The absolute and relative variation of the MS lifetime, $\tau_{\rm H}$, the core helium burning lifetime, $\tau_{\rm He}$ and the BSG lifetime, $\tau_{\rm BSG}$. The individual values of each model are shown in Tables \ref{properties_hedep}.}
	\begin{tabular}{ccccccc}\\
\toprule
			  & $\Delta\tau_{\rm H}$/Myrs & $\delta\tau_{\rm H}$ & $\Delta\tau_{\rm He}$/Myrs & $\delta\tau_{\rm He}$ & $\Delta\tau_{\rm BSG}$/Myrs & $\delta\tau_{\rm BSG}$ \\
			\midrule
			$15\,\Msun$ & 3.35 \textcolor{blue}{(4.13)} & 24.38\% \textcolor{blue}{(30.06\%)} & 0.69 \textcolor{blue}{(0.69)} & 73.40\% \textcolor{blue}{(73.40\%)} & 1.16 \textcolor{blue}{(1.16)} & 5800.0\% \textcolor{blue}{(5800.0\%)} \\[0.5em]
			$20\,\Msun$ & 1.88 & 19.87\% & 0.36 & 50.00\% & 0.88 & 4400.0\% \\[0.5em]
			$25\,\Msun$ & 1.25 & 16.82\% & 0.25 & 40.98\% & 0.54 & 2702.5\% \\
\bottomrule
	\end{tabular}
		\label{relativeVariation_lifetimes}
		\begin{tablenotes}
		\begin{footnotesize}
			\item \textbf{Notes:} See Table \ref{relativeVariation_coremasses} for the calculation of $\Delta Q$ and $\delta Q$.
			\item[*] The blue values in brackets include the models with no CBM.
		\end{footnotesize}
	\end{tablenotes}
	\end{threeparttable}
\end{table*}
The core hydrogen and core helium burning lifetimes in Table \ref{relativeVariation_lifetimes} show a decreasing variation, relative and absolute, with initial mass. CBM mainly influences the burning lifetimes by extending the convective core and providing more fuel for the burning phase. The models with higher initial masses consume their fuel faster, hence, the smaller variation in lifetimes with increasing initial mass. The differences of the core hydrogen burning lifetimes are nearly completely due to the choice of $\fcbm$ and only the models with no CBM show a dependence on the boundary criterion. The relative variation of the helium burning lifetimes is more than twice the relative variation of the hydrogen burning lifetimes. Similar to hydrogen burning, these differences in the helium burning lifetimes is mainly given by the amount of CBM but there is also a small dependence on the boundary criterion. The variation in the BSG lifetimes is extreme but this is to be expected considering the uncertainty connected with this phase. This huge variation translates into the uncertainty of the total mass in Table \ref{relativeVariation_coremasses}.\\ 
The blue values in Tables \ref{relativeVariation_coremasses} and \ref{relativeVariation_lifetimes} represent the same variations but they include the $15\,\Msun$ models with no CBM.  These variations are larger, mainly because of the pure Ledoux model with slow semiconvection, which has much smaller cores (see Table \ref{properties_hedep}).\\

In Sections \ref{HydrogenBurning} and \ref{heliumBurning} we have shown that the helium and carbon-oxygen core masses increase with increasing amount of CBM. Furthermore, CBM enhances the MS width and prevents the occurrence of \textit{convective fingers}. Also, models with more CBM have longer MS and core helium burning lifetimes (Table \ref{properties_hedep}) and experience more mass loss (Fig.~(\ref{MassLossRates_XcHe4_15M_f00p002_variousMixing})). These are effects which are generated by rotation as well \citep[e.g.][]{Heger2000, Meynet2000}. Therefore, some solutions of stellar models might not be singular and care has to be taken when trying to fit 1D stellar evolution models to observations. In this work we studied non-rotating stellar models in order to investigate the effects of CBM without blurring of rotation-driven mixing. In reality, both processes occur and influence each other \citep[e.g.][]{Brun2017, Korre2019} but it is still an open question how convection and rotation interact with each other.\\

\citet{Gabriel2014} discussed the important issue of discontinuities at convective boundaries and how to choose the `right' convective boundary location in the framework of the MLT. We find that similar issues arise in the calculations with no CBM, especially with the \textit{Ledoux} criterion. CBM removes possible discontinuities at the convective boundary. Therefore, the problem with a discontinuity in the chemical composition or its gradient at the convective boundary is reduced, depending on the amount of CBM and amount of resolution at the boundary. In thin convective layers such as the \textit{convective fingers} the problems still arise. However, these convective regions might be a relic of 1D stellar evolution and might not occur in reality (see discussion further down). Nevertheless, in most of the 1D stellar evolution codes the convective boundary is determined before the CBM is applied. Hence, the problem is not solved but rather avoided.\\

The \textit{convective fingers} discussed in Section \ref{HydrogenBurning} do not influence the  simulation sustainably, except if they are able to touch the convective hydrogen core or if they persist until the ICZ appears. Especially the first event can lead to significant changes of e.g. the helium core mass and hydrogen burning lifetime. We demonstrated that there is a limiting amount of CBM above which no \textit{convective fingers} appear. This limit, however, increases with increasing initial mass. On the other hand, observations suggest, that the required amount of CBM during the MS evolution are above the $\fcbm$ limits for \textit{convective fingers}. Therefore, they might not occur at all. If they do occur, there are other issues, such as the high diffusion coefficient and convective velocity predicted by the MLT, which seem unrealistic for thin convective layers. Furthermore, if some sort of mixing above the convective hydrogen core takes place, its nature needs to be determined, i.e. whether it has a finger-like structure as found by e.g. \citet{Langer1985} and discussed in our work or whether it is more a slow constant mixing as suggested by \citet{Schwarzschild1958} and implemented in the MESA code \citep{Paxton2019}. Another possibility could be to limit the mixing efficiency in thin convective layers by the distance to the convective boundary. However, \textit{convective fingers} might be a relic of 1D stellar evolution and finite timestepping, that introduce discontinuities at the location of the convective boundary of the retreating hydrogen core at a certain timestep. Moreover, 3D simulations cleary show that the convective boundary is dynamic and fluctuates \citep[e.g.][]{Cristini2017, Jones2017}, which affects the chemical composition profile that is left behind of the retreating hydrogen core. Furthermore, 3D simulations clearly show that at the interface of the convective and radiative region internal gravity waves are generated, that propagate to the surface \citep{Cristini2017, Edelmann2019}. How much these waves mix the composition needs to be determined but they definitely affect the radiative region above the convective core.\\

Some of the convective helium cores in Fig.~(\ref{convHeliumCoreMass}) show core breathing pulses \citep{Castellani1985}. There is evidence that these \textit{core breathing pulses} might be of a theoretical or numerical nature \citep{Constantino2016, Farmer2016}. We note that the core breathing pulses that occur in some of our models always appear after the ICZ disappeared and the energy generation of the hydrogen shell drops (see Figs.~(\ref{kippenhahn_15M}) and (\ref{luminosities_15M})). At this point the convective core experiences an increase in pressure and readjusts itself. This process, however, is dynamic and time dependent. The pulses of the convective core could therefore be a result of the 1D mixing prescription \citep[e.g.][]{Renzini1988}, hence, they could be the issue of idealised physics rather than a numerical problem. Moreover, we note that when the CBM zone is large enough, this scenario does not occur, i.e. $\fcbm \geq 0.022$ \citep[][found a similar dependence and proposed a entrainment rate for the convective helium core]{Constantino2017}. It might be that \textit{core breathing pulses} occur when the envelope to core ratio is below a critical value.\\

A complete understanding of the blue versus red evolution after the MS is still missing. Several ideas have been suggested \citep[e.g.][]{Renzini1992, Sugimoto2000, Stancliffe2009} but there is no general accepted solution. Nevertheless, it is known that the BSG to RSG ratio depends strongly on internal mixing processes \citep[e.g.][]{Langer1995, Georgy2014, Schootemeijer2019}. There are two possible ways massive stars evolve into the BSG region. Either they evolve from the MS to the BSG region (type \Romannum{1}) or they evolve directly from the MS to the RSG region and then back towards the blue region (type \Romannum{2}). \citet{Ekstroem2012} find in their evolutionary grid that BSG type \Romannum{2} occurs in massive stars of about $20\,\Msun$ or higher; this limit depends on the mass loss rates that are assumed during the RSG phase \citep[e.g.][]{Georgy2012a}. The two types of BSGs have a different mass, structure and surface abundances. Furthermore, their following evolution and supernovae type will differ \citep[e.g.][]{Georgy2012b, Yoon2012, Eldridge2013}.\\
In Section \ref{ICZ} we discussed that BSGs in our grid are either type \Romannum{1}, with small and intermediate values of $\fcbm$, or possibly type \Romannum{2}, when large values of $\fcbm$ are used, depending on the initial mass. This is due to the impact of the ICZ on the evolution of the star. The type \Romannum{1} BSGs models move to the RSG region as soon as the ICZ disappears and the hydrogen shell weakens. Nonetheless, some models spend more than half, or nearly all, of their core helium burning phase as BSGs depending on the convective boundary criterion and the amount of CBM (Table \ref{properties_hedep}). The type \Romannum{2} BSG phase only occurs in the $25\,\Msun$ models with large amounts of CBM and the star only spends a short fraction of core helium burning as this type. Nevertheless, this model becomes a WR star at the end of core helium burning, which will change its further evolution and final fate. Moreover, we find that semiconvection with the efficiencies tested has no remarkable impact on the BSG to RSG ratio. Recently, \citet{Schootemeijer2019} compare a grid of stellar models with varying amounts of internal mixing to observation of massive stars in the Small Magellanic Cloud and conclude that a medium amount of CBM (they use the penetrative `overshoot' with $0.22 \lesssim \alpha_{\rm ov} \lesssim 0.33$) and efficient semiconvection ($\alpha_{\rm sc} \gtrsim 1$) is needed to match observations. Furthermore, they find that inefficient semiconvection ($\alpha_{\rm sc} \lesssim 1$) can be ruled out because not enough BSGs are created. We agree with \citet{Schootemeijer2019} that in general less CBM favours BSGs but a straight comparison is not possible due to the different initial metallicity \citep[at a lower initial metallicity, more BSGs are expected, e.g.][]{Georgy2013}. Nevertheless, we seem to contradict \citet{Schootemeijer2019} in our claim that the relative importance of semiconvective mixing decreases with increasing amount of CBM. The simple explanation is the fact that \citet{Schootemeijer2019} only apply CBM at the convective hydrogen core whereas we include it at all convective boundaries. This leads to a different behaviour of the ICZ, hence, red-wards evolution (see Section \ref{ICZ}). Furthermore, \citet{Schootemeijer2019} apply the penetrative CBM scheme (`step-overshoot') which creates a discontinuity in the chemical composition at the boundary. We conclude that in our simulations a more efficient semiconvection ($\alpha_{\rm sc} > 1$) might only affect the outcome in the calculation with no or a small amount of CBM. \\
Another important impact of possible blue loops is the amount of mass loss during the RSG phase, especially at solar metallicity. This might strip the star of its envelope and it moves back to the BSG region. Indeed, if we enhance the mass loss rates due to dust formation during the RSG phase following the prescription of \citet{vanLoon2005}, the models with a fast red-wards evolution go back into the blue region. The models that stay in the blue region right after the MS do not experience enough mass loss once they enter the RSG phase to do a blue-loop. However, the mass loss rates during the RSG branch are very uncertain and it is not sure, which mass loss prescription is the correct one, if any. \citet{Saio2013} suggest to distinguish between the two BSG types with radial pulsations \citep[e.g.][]{Bowman2019} in addition to the CNO surface enrichment, i.e. BSG type \Romannum{2} exhibit radial pulsations. Observed number ratios of BSG type \Romannum{1} or \Romannum{2} to RSG would help to constrain the internal mixing process of e.g. the ICZ and the mass loss rates.\\
\citet{Vink2010} find a steep drop in the rotation rates of B supergiants and propose two possible explanations for their nature. We want to discuss the possibility that this could be a consequence of the ICZ. As discussed in Section \ref{BSG_vs_RSG}, if a massive star has a strong ICZ it can spend $75 - 90\%$ of its helium burning phase as a BSG right after the MS. On the other hand, rotation tends to smooth the structure above the convective hydrogen core, which reduces the ICZ and leads to a faster red-ward evolution \citep{Maeder2001}. Thus, it is much less probable for these stars to be observed during the crossing to the RSG branch. Therefore, the observed BSG after the MS are most probably the ones with a strong ICZ, hence, no or a slow rotation. If this is indeed the case, such observations could be used to constrain the mixing of the ICZ.\\

For simplicity, we use the same $\fcbm$ at all convective boundaries in all evolutionary stages, core and shell convective zones and all initial masses in our simulations. However, the amount of CBM depends on the stiffness of the boundary \citep[e.g. bulk Richardson number,][]{Cristini2019} and different $\fcbm$s might be needed for different stellar stages, resulting in different evolutionary paths. It is an ongoing effort to create a CBM prescription, which depends on the physics of the boundary rather than the parametrisation \citep[e.g.][]{Pratt2017, Arnett2018b, Arnett2019a}.\\

In Section \ref{physicalIngredients} we mentioned the use of \texttt{MESA}s \texttt{MLT++} in models that experience envelope inflation. Only models with large amounts of CBM experience density and gas-pressure inversions in their outer layers. Recent efforts to constrain internal mixing in massive stars with observations \citep[e.g.][]{Brott2011, Castro2014, Schootemeijer2019, Higgins2019} indicate that stars in the mass range studied here have larger amounts of CBM than often assumed in `\textit{traditional state-of-the-art}' stellar evolution models. Therefore, models in the mass range $15$ - $25\,\Msun$ might experience inflated envelopes. However, the stability and treatment of such radiation-dominated envelopes is still an open question \citep[e.g.][]{Joss1973, Maeder1987, Langer1997, Bisnovatyi-Kogan1999, Maeder2009, Suarez-Madrigal2013} and is yet another uncertainty in stellar evolution, which affects the post-MS evolution. However, the transport of energy by the classical treatment of MLT in such regions is clearly out of its applicability range.

\section{Conclusions} \label{conclusions}

We calculated two grids of stellar models, each with the \textit{Ledoux} and the \textit{Schwarzschild} convective boundary criterion, for the three initial masses $15$, $20$ and $25\,\Msun$, in order to investigate the impact of some CBM uncertainties. In each grid we varied the amount of CBM between $0.004$ and $0.05$ and, in the \textit{Ledoux} case, the semiconvective efficiency. In Sections \ref{HydrogenBurning}, \ref{ICZ} and \ref{heliumBurning} we presented the impact of the uncertainties on the stellar structure. Our findings summarise as follow:
\begin{enumerate}
	\item During the MS evolution the difference of the convective core size due to the two convective stability criteria converges in all models when CBM is included. Furthermore, the region above the core converges with more CBM. We find that the minimal amount of CBM above where no \textit{convective fingers} are present increases with initial mass. This indicates that \textit{convective fingers} might be a relic of 1D stellar evolution since observations suggest larger amounts of CBM on during the MS evolution in massive stars.
	\item The width of the MS broadens significantly with increasing amount of CBM and with the largest $\fcbm$ value the terminal-age main-sequence bends slightly to cooler effective temperatures with increasing initial mass, which is more in agreement with recent observations \citep[e.g.][]{Castro2014, McEvoy2015}. The width of the MS is nearly independent of the convective boundary criterion and semiconvective efficiency.
	\item The initial location of the ICZ strongly depends on the stability criterion, regardless of the amount of CBM. The \textit{Schwarzschild} criterion predicts an overlap with the hydrogen burning shell, whereas the \textit{Ledoux} criterion not. The further evolution of the ICZ is largely determined by the amount of CBM. More mixing shortens the lifetime of the ICZ and leads to an overlap with the hydrogen burning shell in the Ledoux models. An overlap between the two boosts the latter, leading to crucial differences in the further central and surface evolution of the model.
	\item The relative importance of semiconvection drastically decreases with an increasing amount of CBM.
	\item Generally, more CBM leads to larger core masses and longer lifetimes. Models with large amounts of CBM behave more like models of a higher initial mass but less CBM in terms of core masses, core helium burning lifetimes and nucleosynthesis. This would lead to a different further evolution, SN progenitor structure and explodability than currently presented in the literature.
\end{enumerate}
In Section \ref{nucleosynthesis_heburn} we showed the impact of the CBM uncertainties on the nucleosynthesis during central helium burning. In the $15\,\Msun$ models the $^{12}$C to $^{16}$O ratio decreases as $\fcbm$ increases due to the larger amount of fuel available during the late stage of this burning phase. The $^{12}$C to $^{16}$O ratio is naturally saturated in the models with higher initial masses due to the activation of other particle capture reactions. Furthermore, we find an increase of the weak s-process activity in the simulations with larger amounts of CBM. This might affect the peak production of the weak s-process and will be subject to further studies.\\
In Section \ref{BSG_vs_RSG} we discussed the impact of the ICZ on the surface evolution of the star. The simulations that predict a strong ICZ remain in the BSG region until the convective shell recedes, whereas models with a short ICZ move directly to the RSG branch. As a result, some models spend nearly their whole core helium burning lifetime as BSG, depending on the strength of the ICZ. On the other hand, some of the more massive models very quickly enter the RSG phase after the MS and become BSG, and later on WR stars, at the end of core helium burning due to strong mass-loss. This not only affects the BSG to RSG ratio but also the total mass at core helium depletion, the further evolution, the pre-SN structure and explodability of these models.\\
In Tables \ref{relativeVariation_coremasses} and \ref{relativeVariation_lifetimes} we presented the absolute and relative variations of the total mass, the core masses and the stellar life times to show which part of stellar evolution is mostly affected by the uncertainties of CBM. The strongest affected values are the BSG lifetimes and, correlated, the mass loss rates. The importance of the latter increases with initial mass. The core masses show an uncertainty of $\sim 40\%$ for the helium core mass and $\sim 50 - 70\%$ for the carbon-oxygen core mass. The lifetimes show a relative variation of $\sim 15-25\%$ for the hydrogen burning phase and $\sim 40 - 70\%$ for the helium burning lifetime. The biggest uncertainty for all phases comes from the amount of CBM. The choice of the boundary criterion, either \textit{Ledoux} or \textit{Schwarzschild}, mainly influences the ICZ, by determining its initial location, and the growth of the convective helium core. The convective cores, however, grow to similar maximal sizes and the difference introduced by the boundary criterion is small. Therefore, the choice of the boundary criterion has nearly no impact on the MS evolution but it crucially affects the surface evolution.\\
This work shows the need to improve the treatment of convective boundaries in 1D stellar evolution codes in order to have more reliable predictions of the evolution of (massive) stars. The ICZ, for example, should be investigated in multi-dimensional simulations to constrain the amount of mixing at the convective boundary and to test the two boundary criteria. Furthermore, observations of the ratio of the two BSG types \citep[i.e. a BSG right after the MS or after the RSG phase, e.g.][]{Saio2013} and RSGs would help to constrain the internal mixing processes and the boundary criterion. Also, asteroseismic observations may help in constraining the amount of extra mixing at the convective boundary. However, constraining the amount of mixing is only a first step. In order to have more reliable predictions ultimately a non-parametrised but physical theory is sought, which is work in progress by many teams.

\section*{Acknowledgements}

We are grateful to the anonymous referee who helped to increase the quality of this manuscript. EK thanks M.Pignatari and O.Clarkson for the valuable discussions. This article is based upon work from the ``ChETEC'' COST Action (CA16117), supported by COST (European Cooperation in Science and Technology. This work benefited from discussions at the NuGrid/JINA/ChETEC School: Software Tools for Simulations in Nuclear Astrophysics, supported by the National Science Foundation under Grant No. PHY-1430152 (JINA Center for the Evolution of the Elements). The authors acknowledge support from the IReNA AccelNet Network of Networks, supported by the National Science Foundation under Grant No. OISE-1927130. RH acknowledges support from the World Premier International Research Centre Initiative (WPI Initiative). CG has received funding from the European Research Council (ERC) under the European Union's Horizon 2020 research and innovation program (grant agreement No 833925). This research has made use of the NASA's Astrophysics Data System Bibliographic Services.\\

\textit{Software}: \texttt{MESA} (\href{http://mesa.sourceforge.net}{https://reaclib.jinaweb.org}), \texttt{Python} (\href{https://www.python.org}{https://www.python.org}), \texttt{matplotlib} \citep{Hunter2007}, \texttt{NumPy} \citep{Walt2011}, \href{https://nugrid.github.io}{NuGrid} tools, JINA reaclib database (\href{https://reaclib.jinaweb.org}{https://reaclib.jinaweb.org})

\bibliographystyle{mnras}
\bibliography{publicConvection.bib}

\bsp
\label{lastpage}
\end{document}